\documentclass[11pt]{article}

\usepackage[many]{tcolorbox}
\usepackage{booktabs}

\usepackage{inputenc}
\usepackage{fullpage}
\usepackage[letterpaper, portrait, margin=1.in]{geometry}
\usepackage{amsmath}
\usepackage{amssymb}
\usepackage{textgreek}
\usepackage{xcolor}
\usepackage{inconsolata}
\usepackage[protrusion=true,expansion=true]{microtype}
\usepackage[noblocks]{authblk}
\usepackage[sorting=none,style=numeric-comp]{biblatex}
\usepackage[hyperfootnotes=false]{hyperref}
\usepackage[multiple]{footmisc}
\usepackage{graphicx}
\usepackage{csquotes}    
\usepackage{enumitem}
\usepackage{dirtytalk}
\usepackage{epigraph}
\usepackage{etoolbox}
\usepackage{tabularx}
\usepackage{wasysym}
\usepackage{gensymb}
\usepackage[firstpage=true]{background}
\usepackage{titlesec}

\newcommand\snowmass{\begin{center}\rule[-0.2in]{\hsize}{0.01in}\\\rule{\hsize}{0.01in}\\
\vskip 0.1in Submitted to the  Proceedings of the US Community Study\\ 
on the Future of Particle Physics (Snowmass 2021)\\ 
\rule{\hsize}{0.01in}\\\rule[+0.2in]{\hsize}{0.01in} \end{center}}

\backgroundsetup{contents={\parbox{6.5in}{\snowmass}}, scale=1,placement=top,opacity=1,color=black,position={3.25in,1.2in}}

\titleformat{\paragraph}
{\normalfont\normalsize\bfseries}{\theparagraph}{1em}{}
\titlespacing*{\paragraph}
{0pt}{3.25ex plus 1ex minus .2ex}{1.5ex plus .2ex}

\newcommand{\randd}{R$\&$D }
\newcommand{\nbsn}{Nb$_3$Sn }
\newcommand{\ybco}{\ensuremath{\mathrm{YBa_2Cu_3O_{7-\delta}}}}
\newcommand{\bi}{\ensuremath{\mathrm{Bi_2Sr_2CaCu_2O_x}}}
\newcommand{\bit}{\ensuremath{\mathrm{Bi_2Sr_2Ca_2Cu_3O_{10+x}}}}

\def\beq{\begin{equation}}
\def\eeq#1{\label{#1}\end{equation}}
\def\eeqn{\end{equation}}

\newenvironment{Eqnarray}%
   {\arraycolsep 0.14em\begin{eqnarray}}{\end{eqnarray}}
\def\beqa{\begin{Eqnarray}}
\def\eeqa#1{\label{#1}\end{Eqnarray}}
\def\eeqan{\end{Eqnarray}}

\let\bar=\overbar

\def\lsim{\mathrel{\raise.3ex\hbox{$<$\kern-.75em\lower1ex\hbox{$\sim$}}}}
\def\gsim{\mathrel{\raise.3ex\hbox{$>$\kern-.75em\lower1ex\hbox{$\sim$}}}}

\def\del{\partial}
\def\Dslash{\not{\hbox{\kern-4pt $D$}}}
\def\dslash{\not{\hbox{\kern-2pt $\del$}}}
\def\pslash{\not{\hbox{\kern-2pt $p$}}}
\def\ETmiss{\not{\hbox{\kern-4pt $E$}}_T}

\def\Dlr{\mathrel{\raise1.5ex\hbox{$\leftrightarrow$\kern-1em\lower1.5ex\hbox{$D$}}}}

\def\MSB{{\bar{M \kern -2pt S}}}
\def\msb{{\bar{\scriptsize M \kern -1pt S}}}

\def\drb{{\bar{\scriptsize D \kern -1pt R}}}

\title{A Strategic Approach to Advance Magnet Technology for  Next Generation Colliders}

\author[ ]{Authors (alphabetical):} 
\author[2]{G. Ambrosio}
\author[3]{K. Amm}
\author[3]{M. Anerella}
\author[2]{G. Apollinari}
\author[1]{D. Arbelaez}
\author[9]{B. Auchmann}
\author[4]{S. Balachandran}
\author[2]{M. Baldini}
\author[6]{A. Ballarino}
\author[4]{S. Barua}
\author[2]{E. Barzi}
\author[1]{A. Baskys}
\author[1]{C. Bird}
\author[1]{J. Boerme}
\author[4]{E. Bosque}
\author[1]{L. Brouwer}
\author[1]{S. Caspi}
\author[4]{N. Cheggour}
\author[2]{G. Chlachidze}
\author[4]{L. Cooley}
\author[4]{D. Davis}
\author[1]{D. Dietderich}
\author[2]{J. DiMarco}
\author[4]{L. English}
\author[1]{L. Garcia Fajardo}
\author[1]{J.L. Rudeiros Fernandez}
\author[1]{P. Ferracin}
\author[1]{S. Gourlay}
\author[3]{R. Gupta}
\author[1]{A. Hafalia}
\author[4]{E. Hellstrom}
\author[1]{H. Higley}
\author[4]{I. Hossain}
\author[8]{M. Jewell}
\author[4]{J. Jiang}
\author[1]{GM. Juchno}
\author[4]{F. Kametani}
\author[2]{V. Kashikhin}
\author[2]{S. Krave}
\author[3]{M. Kumar}
\author[3]{F. Kurian}
\author[7]{A. Lankford}
\author[4]{D. Larbalestier}
\author[4]{P. Lee}
\author[1]{G. S. Lee}
\author[2]{V. Lombardo}
\author[1]{M. Marchevsky}
\author[2]{V. Marinozzi}
\author[1]{C. Messe}
\author[10]{J. Minervini}
\author[1]{C. Myers}
\author[1]{M. Naus}
\author[2]{I. Novitski}
\author[5]{T. Ogitsu}
\author[3]{M. Palmer}
\author[1]{I. Pong}
\author[1]{S. Prestemon}
\author[3]{C. Runyan}
\author[1]{G.L. Sabbi}
\author[1]{T. Shen}
\author[2]{S. Stoynev}
\author[2]{T. Strauss}
\author[4]{C. Tarantini}
\author[1]{R. Teyber}
\author[4]{U. Trociewitz}
\author[1]{M. Turqueti}
\author[2]{M. Turenne}
\author[2]{D. Turrioni}
\author[1]{G. Vallone}
\author[2]{G. Velev}
\author[1,11]{S. Viarengo}
\author[1]{L. Wang}
\author[1]{X. Wang}
\author[2]{X. Xu}
\author[5,6]{A. Yamamoto}
\author[1]{S. Yin}
\author[2]{A. Zlobin}

\affil[1]{Lawrence Berkeley National Laboratory, Berkeley, CA 94720, USA}
\affil[2]{Fermi National Accelerator Laboratory, Batavia IL 60510-5011, USA}
\affil[3]{Brookhaven National Laboratory, Upton, NY 11973-5000, USA}
\affil[4]{ASC / NHMFL / Florida State University, Tallahassee, FL 32310, USA}
\affil[5]{KEK, Tsukuba, Ibaraki, Japan}
\affil[6]{CERN, Geneva, Switzerland}
\affil[7]{University of California, Irvine, CA  92697-4575}
\affil[8]{University of Wisconsin-Eau Claire, Eau Claire, WI 54702-4004 }
\affil[9] {Paul Scherrer Insititue, Villigen, Switzerland}
\affil[10] {Massachusetts Institute of Technology, Cambridge, Ma.}
\affil[11] {Politecnico di Torino, Torino, Italy}

\date{Accelerator Frontier (AF), Multi-TeV Colliders (AF4)}

\DeclareUnicodeCharacter{2212}{-}
\begin{document}
\maketitle


\newpage

\tableofcontents

\newpage


\newcounter{statements}
\setcounter{statements}{1}

\section{Executive Summary}
As the High Energy Physics community considers future discovery machines based on circular colliders, the performance and cost drivers for such facilities are of fundamental importance. The leading consideration for such a facility is the magnet technology which steers and focuses the particle beams.  As a result of the last Snowmass and P5 process (2013-2014), DOE-OHEP initiated the US Magnet Development Program (MDP), a general \randd program that pulls together longstanding individual HEP research groups at DOE laboratories and University programs focused on magnet technology under a common collaboration, with focused mission and goals and constructive review and guidance from a Technical Oversight Committee.

The general magnet \randd pursued by the MDP focuses on advancing magnet technology irrespective of a specific project, and hence provides broad and lasting value to the community as it explores future hadron colliders as well as muon colliders. The combination of clarity in mission and goals, coupled with the transparent, collaboration-friendly research paradigm of the MDP, has led to internationally recognized leadership in the field, and to growing synergies with the NSF-funded NHMFL; with other DOE offices, in particular DOE-OFES; and with industry, for example through the SBIR program. These synergies significantly enhance the effectiveness of the MDP. Furthermore, international collaborations exist on specific technical topics, and we seek to enhance collaborations with our European counterparts now that the European Strategy document has been issued and the associated technology roadmaps have been developed. 

To explore long range opportunities for future HEP colliders, the MDP focuses on high field magnet technology. All advanced superconductors of relevance to HEP high field magnets, in particular the low-temperature superconductor \nbsn and the high temperature superconductors REBCO and Bi2212, are strain sensitive and brittle, characteristics that challenge magnet design, fabrication, and operational performance.  The MDP is exploring and developing \say{stress-managed} design concepts to enable the use of these materials in high field accelerator magnet configurations. We note that the magnet designs have relevance to both hadron and muon collider needs. For muon colliders, there is furthermore need to develop very high field solenoids for muon production and for muon beam cooling. Expertise from MDP can be applied in these areas as well, and we see strong synergies in that arena with ongoing development in high field magnets for materials research, exemplified by the record solenoid field developed and produced by the National High Magnetic Field Laboratory (NHMFL), as well as advances in REBCO magnet technology by DOE-OFES and  by private industry in the nascent - but rapidly developing - compact fusion realm. 

A major element of the MDP is the development of technologies critical to the understanding of magnet performance; these include advanced modeling capabilities, the development and implementation of novel diagnostics, and the exploration of new concepts that can impact the rate at which accelerator magnets \say{train} up to full field. Advances in these areas provide lasting benefit for HEP, and are highly valued by the broader superconducting magnet community, both within the DOE Office of Science, and also in industry.

At the heart of superconducting magnet technology are the superconductors themselves; the US has longstanding leadership in collaborating with industry to develop new and improved superconductors, and it is critical that the successful partnership be further grown and nurtured to support HEP's future needs. HEP is working closely with the new Accelerator \randd and Production Office (ARDAP) to develop a strategic plan that can strengthen US industry in this arena and support HEP's long term needs for conductor performance and cost-effective conductor production.

General long range magnet \randd is a critical foundation for HEP, providing a strong basis to develop strategies for future colliders, growing the next generation of scientists and engineers in the field of accelerator magnets, so as to provide the foundation for directed-\randd that will ultimately be required to scale up and finalize magnet designs for a future collider project. Currently the MDP is focused on exploring stress-managed structures and in testing hybrid HTS/LTS accelerator magnets as a cost-effective means of achieving high-field accelerator magnets; in parallel, designs are underway to explore 20 T and beyond, which may be within reach in the next decade. 

The HEP MDP collaboration is now well established, with a record of achievements and a clear understanding of the opportunities and challenges that lay ahead. General magnet \randd, when well coordinated and managed and fully leveraging synergistic activities, is an excellent investment for high energy physics. The experience from MDP indicates that the scale of investment needed to aggressively pursue magnet \randd with the goal of enabling a foundation for a future readiness program, and ultimately collider project,  is $\sim\$15M$ per year, with an additional $\sim \$20$\% (i.e. $\$3$M) in conductor procurement from industry to support the magnet development needs. In summary,  a long range magnet \randd program, designed to advance magnet technology while fully leveraging the broader community's strengths, is vital to HEP and the future of particle physics.   

\newpage

\section{Introduction}
Today’s colliders are built on a foundation of superconducting magnet technology that provides strong dipole magnets to maintain the beam orbit and strong focusing magnets to enable the extraordinary luminosity required to probe physics at the energy frontier. The dipole magnet strength plays a critical role in dictating the energy reach of a collider \cite{Skandszn}, and the superconducting magnets are arguably the dominant cost driver for future collider facilities. As the community considers opportunities to explore new energy frontiers, the importance of advanced magnet technology – both in terms of magnet performance and in the magnet technology’s potential for cost reduction – is evident, as the technology status is essential for informed decisions on targets for physics reach and facility feasibility. 

\section{Magnet needs for future colliders}
\subsection{The energy frontier landscape - facilities under consideration}
There are a large number of facilities under consideration to probe new realms of the energy frontier, i.e. \say{discovery machines}, as well as a variety of facilities focused on high beam intensity, i.e. machines focused on probing known energy ranges with significant improvements in statistics to thoroughly explore physics phenomena and identify rare events \cite{Benedikt1,Shiltsev2}. Here we consider those facilities that are ring-based, i.e. that seek to leverage recirculating \say{stored} beams as a cost-effective means of achieving high energies and high luminosities. Assuming the counter-rotating beams are at the same energy $E$, the resulting center of mass energy is $E_{c.m.}=\sqrt{s}=2E$. Central to ring-based accelerators are the magnets that bend the beams, as well as the radio-frequency cavities that provide increase energy, and that maintain beam energy to compensate for energy loss due to synchrotron radiation. The bend radius $\rho$ for a beam of charge $e$ and momentum $p=mv$ is simply $p=eB\rho$. Hence higher magnetic field enables proportionally higher energies for the same ring radius. For any future high energy collider, magnet technology arguably drives both the energy reach and the cost of the facility. 

Discovery machines have historically used hadron beams, e.g.  protons. The world's premier energy frontier facility, the Large Hadron Collider at CERN, currently achieves a center of mass energy $E_{c.m.} \sim 13$ TeV, colliding two counter-rotating proton beams of energy $6.5$ TeV. A natural approach to reach new physics is to significantly enhance the beam energy and resulting $E_{c.m.}$. The most aggressive example is the proposed "Future Circular Collider" (FCC), a facility advocated by CERN that would target $E_{c.m.}\sim100$ TeV \cite{Benedikt2} in a new tunnel of circumference $\sim 100$ km. A competing proposal, the Super proton-proton Collider (SppC) proposed by China, targets an energy of $E_{c.m.}\sim 80$ TeV \cite{Gao2}. These machines are designed to access new physics, in particular physics beyond the Standard Model, that promise answers to questions related, for example, to the nature of dark matter, the total absence of antimatter in the universe, and the peculiarities of neutrinos. 

An alternative approach to high energies is to leverage the large mass of muons, while benefitting from their point-particle-like nature, i.e. Lepton physics. The muon mass $m_\mu$ is $207$ times that of the electron; since synchrotron radiation scales with $m^{-4}$, it is in principle feasible to consider a ring-based muon collider at the energy frontier. Furthermore, for new physics a hadron interaction requires $\sim 7$ times the $E_{c.m.}$ energy; for example, a $\sqrt{s}=14$ TeV muon collider is competitive with a $\sqrt{s}=100$ TeV proton collider.

From an accelerator technology point of view, a high energy hadron collider, although technically very challenging, is arguably well understood. A muon collider, however, has unique challenges that must be understood and overcome - these relate primarily to the production and control of the very short-lived muon beams.  

\subsection{Hadron colliders - impact of magnet technology developments}
\begin{figure}[h!]
\centering
        \includegraphics[scale =0.7, angle=0]{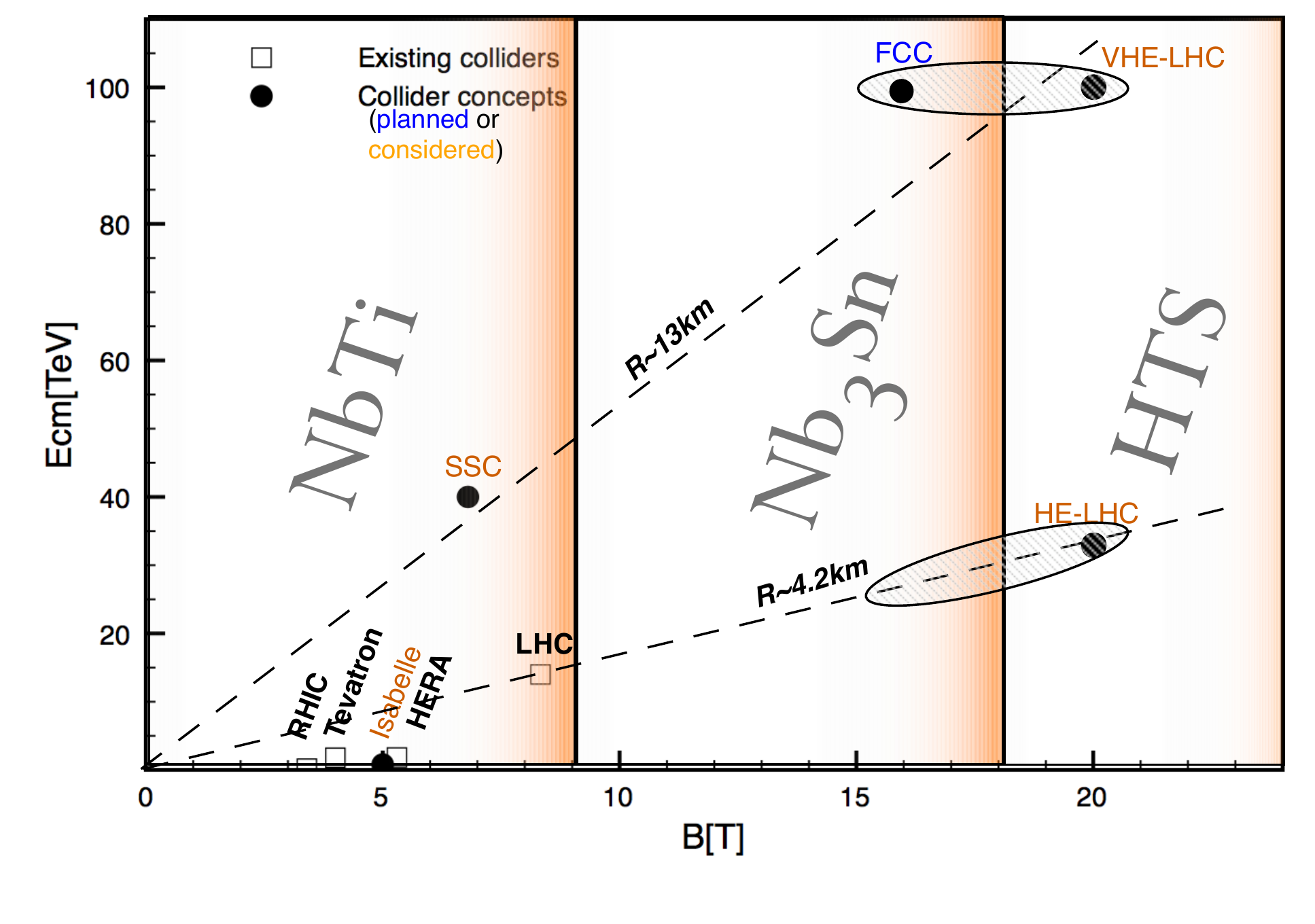}
       \caption{The center of mass energy for a hadron collider as a function of the dipole magnet field strength, assuming the $E_{c.m.}=0.3B[T]R[km]$ scaling, consistent with reasonable filling-factors of magnets on the ring circumference.}
       \label{fig:field-regimes}
\end{figure}

For the main ring magnets, i.e. dipole magnets, of a hadron collider, the linear relationship between beam energy, ring radius, and magnetic field strength provide the basis for facility optimization for physics reach and for cost. For site-limited facilities, e.g. when an existing tunnel will be re-used, the magnet technology  dictates the physics reach that can be obtained. For a \say{green-field} facility, i.e. where a new tunnel and new magnets are being considered, both the magnet technology and the ring circumference are parameters that impact the physics reach; a reasonable estimate for the beam energy is $E=0.3B R$, with energy $E$ in TeV, field $B$ in T, and radius $R$ in km. The cost of the tunnel is arguable $\propto R$; the cost of magnets is typically thought to scale $\propto B$, but we must recognize that the required  conductor mass scales $\propto B^2$, assuming the average current density is roughly constant independent of field - this assumption appears to be reasonably valid to-date. Furthermore, the scaling relation is significantly impacted as the field strength approaches a superconductor's upper critical field $B_{c2}$. Figure \ref{fig:field-regimes} shows schematically the transition from one superconductor to another as a function of field, and its relation to hadron collider center of mass energy reach.  

\subsection{Muon colliders - impact of magnet technology developments}
The driving considerations for a muon collider relate to a) the short lifetime of the muon ($\sim 2.2\mu$ s at rest), and b) the fact that muon production, typically done by colliding a high energy proton beam on a  target (resulting in the production of pions which decay to muons) results in a large beam emittance that requires \say{cooling} to achieve the beam properties needed for high luminosity collisions \cite{Bogomilov1}. 

The efficiency of muon production, via pion production at the target, requires very strong solenoidal magnetic focusing field to maximize the number of pions captured and to focus them as they are accelerated. The collision process furthermore results in significant radiation by-products, further challenging the requirements on the magnetic field system. 

The short muon lifetime  motivates novel magnet and optics schemes to enable very rapid acceleration of the muons from target production to the storage ring where collisions occur. Examples include very fast ramping magnets in a synchronous multi-pass configuration, for example using high-temperature superconductors in novel configurations \cite{Piekarz}, and/or large momentum acceptance magnet optics, such as fixed-field alternating gradient ("FFAG") designs that enable the beam to be accelerated via multiple passes with fixed magnetic field \cite{berg1}. Finally, since even at relativistic velocities the lifetime of the muon is limited, luminosity considerations motivate the most compact storage ring possible, i.e. the highest field dipoles possible to minimize the ring circumference.

The emittance consideration implies the need for significant beam cooling; the most promising technique to date is \say{ionization} cooling, where the muon beam is focused using strong solenoidal fields and passed through material, i.e. \say{absorber}, resulting in (both transverse and longitudinal) momentum reduction; the longitudinal momentum is then restored via (longitudinal) RF electric fields \cite{Bogomilov1}. Since the beam focusing drives the cooling rate and scales with the magnetic field strength, very strong solenoidal fields are required. We note that the large beam emittance also results in a large beam envelope, which implies the need for relatively large apertures in the bending dipoles of the high-energy acceleration stage. 

The collider ring will require high field dipoles to maximize luminosity; for example, at $2$ TeV the muon lifetime, though significantly extended, is only $0.044$ s; the ring circumference then limits the number of turns before decay and hence dictates the interaction opportunities. Furthermore, the collider  ring will  need large aperture high-field magnets to accommodate absorbers to  dissipate  the beam losses, estimated to be at the level of $\sim 500$ W/m.

We note that the solenoid fields required by a muon collider have some similarities with magnet systems being developed for other applications. As examples, the high field solenoids being developed at the NHMFL for condensed matter experiments, including high-field NMR, have many similarities to the solenoid technology needed for muon cooling channels. The larger bore high field solenoid needed for the muon collider target has many commonalities with a central solenoid for a compact fusion machine, and/or with hybrid resistive/superconducting solenoids similar to the NHMFL's 45 T flagship facility, or the more recent series-connected hybrid solenoids built for high field NMR.

Leveraging expertise and investments from other programs with common interests leads to synergies that can significantly speed up magnet development and mitigate risk for future projects.

\begin{tcolorbox}[colback=yellow!10!white,
                     colframe=black!75,
                     coltitle=black,
                     colbacktitle=black!10,
                     fonttitle=\bfseries,
                     title=Statement \thestatements : High field magnets are critical to energy-frontier circular colliders,  
                     center, 
                     valign=top, 
                     halign=left,
                     before skip=0.8cm, 
                     after skip=1.2cm,
                     center title, 
                     width=15cm]
 
     Advanced magnet technology is the driving technology for any energy-frontier circular collider envisioned by the community, fundamentally impacting both science reach and facility cost.
  \end{tcolorbox}
  \stepcounter{statements}

\begin{tcolorbox}[colback=yellow!10!white,
                     colframe=black!75,
                     coltitle=black,
                     colbacktitle=black!10,
                     fonttitle=\bfseries,
                     title=Statement \thestatements : Muon colliders have additional magnet challenges that should be addressed by HEP magnet \randd in synergy with other DOE and NSF programs,  
                     center, 
                     valign=top, 
                     halign=left,
                     before skip=0.8cm, 
                     after skip=1.2cm,
                     center title, 
                     width=15cm]
 There are significant accelerator magnet development efforts that are common to any future circular collider and benefit from a general \randd program. Furthermore, in particular for a muon collider, there are significant synergies with other DOE programs and with industry that should be leveraged to more rapidly and efficiently develop critical magnet technology for HEP.
  \end{tcolorbox}
  \stepcounter{statements}

\section{Status of magnet R\&D and future directions}
The dominant, commercially available conductors today are the low-temperature superconductors (\say{LTS}) NbTi and Nb$_3$Sn, and a suite of high temperature superconductors (\say{HTS}) based on Bismuth Strontium oxides and on rare-earth Barium Copper oxides (\say{REBCO}). We note that all superconducting colliders to-date use the "workhorse" superconductor NbTi; the LHC luminosity upgrade, \say{HiLumi}, will include new interaction region focusing magnets that utilize - for the first time in a collider - the superconductor Nb$_3$Sn.

The HTS materials lack the maturity of their LTS counterparts, both in terms of the scale of industrial production and uniformity and control of properties, as well as in their application to magnets. Nevertheless they are progressing rapidly, due both to potential applications at higher temperature, and to the fact that they continue to carry significant transport current at high field, well beyond that of their LTS counterparts, thereby promising access to higher field superconducting magnets. The  Bi family has two commercial forms: 
\begin{itemize}
    \item Bi$2223$ (\bit), a multi-filamentary superconductor in tape form that does not need further processing. A single industrial supplier produces the tape. To date the material has had only limited use in accelerator magnet applications due to its performance characteristics as compared to other HTS materials.
    \item Bi$2212$ (\bi), a multi-filamentary round wire that requires a high temperature reaction in an Oxygen environment to create the superconducting state. The round wire is amenable to Rutherford cabling, the primary scalable cable architecture used to-date in all HEP colliders.
\end{itemize}
The REBCO family of superconductors are all produced in tape form, and do not require further processing. However, they are not multifilamentary. The primary rare-earth used is Yttrium (i.e. \ybco). 

An important characteristic of all advanced superconductors, including Nb$_3$Sn and the HTS conductors Bi2212 and REBCO, is that they are strain-sensitive. The strain sensitivity manifests itself at low strain levels as a reversible, intrinsic property, i.e. the critical current behaves as $J_c=J_c(B,T,\epsilon)$. At larger strain levels the materials exhibit irreversible reduction in transport current, i.e. \say{degradation}. Important goals in high field magnet research include properly accounting for reversible strain on the performance behavior of high field accelerator magnets, and clarifying the acceptable strain state of the superconductor so as to design magnets that do not exhibit degradation, while maximizing the field producing potential of the superconductor.
\begin{tcolorbox}[colback=yellow!10!white,
                     colframe=black!75,
                     coltitle=black,
                     colbacktitle=black!10,
                     fonttitle=\bfseries,
                     title=Statement \thestatements : Advanced superconductors are strain-sensitive,  
                     center, 
                     valign=top, 
                     halign=left,
                     before skip=0.8cm, 
                     after skip=1.2cm,
                     center title, 
                     width=15cm]
     High field accelerator magnets must be designed to accommodate the strain limitations intrinsic to the superconductors. The strain limits must be thoroughly studied, understood, and quantified so that magnet designs avoid degradation while maximally leveraging the superconductors properties.
  \end{tcolorbox}
  \stepcounter{statements}
  
  HEP has long served as the driving force behind advances in commercial superconductors, and in some cases those commercial products have enabled new commercial applications to flourish - the development of more advanced NbTi wire for the Tevatron, for example, provide critical performance that enabled the - now ubiquitous - Magnetic Resonance Imaging (MRI) business to thrive. Important developments in superconductors continue, both in \nbsn and in the HTS materials; an excellent perspective on developments in that arena, as well as challenges and possible avenues to strengthen and accelerate US capabilities in commercial superconducting materials, is provided in a complementary whitepaper \cite{cooley1}.
 
 \begin{tcolorbox}[colback=yellow!10!white,
                     colframe=black!75,
                     coltitle=black,
                     colbacktitle=black!10,
                     fonttitle=\bfseries,
                     title=Statement \thestatements : HEP leadership in driving advances in commecial superconductors will require strategic planning and investments,  
                     center, 
                     valign=top, 
                     halign=left,
                     before skip=0.8cm, 
                     after skip=1.2cm,
                     center title, 
                     width=15cm]
     The properties of the superconductors used in high field accelerator magnets are critical to the magnet performance, and are a major element in the magnet cost. The ability to continue to innovate new superconductor architectures, and to bring new conductors to industrial maturity for projects, is a key component of magnet development. Nurturing and maintaining a vibrant, competitive ecosystem for superconductor development as well as for reliable, high quality production capability, will require long term strategic planning and commitment from DOE, in close collaboration with industry. 
  \end{tcolorbox}
  \stepcounter{statements}


\subsection{Results in high-field superconducting magnet R\&D }
The DOE office of High Energy Physics has sponsored research into high field accelerator magnets at a number of laboratories over the years, most notably LBNL, FNAL, and BNL, as well as programs at multiple Universities, including most significantly FSU, TAMU, and OSU. The programs have probed a variety of accelerator magnet concepts over the last three decades, with record dipole fields in multiple configurations. As examples:
\begin{itemize}
    \item The \say{Cosine-Theta} approach, used in all colliders to-date. The \say{D20} cosine-theta magnet \cite{D20} was a flagship 4-layer magnet built in the late 1990's that achieved a peak field of $13.5$ T at $1.9$ K. Most recently, the MDPCT1 magnet, built at FNAL within the MDP program (see section \ref{section:MDP}), achieved the record field of $14.5$ T at $1.9$ K in a $50$ mm bore, with excellent field quality \cite{Turrioni0r9}. 
    \item The \say{Common Coil} magnet configuration, wherein racetrack coils are energized in \say{reverse polarity} so as to create a strong magnetic field between the coils, yielding effectively twin apertures. Examples include the \say{RD3c} magnet built by LBNL \cite{Scanlan77}, the  HFDC01 magnet developed by FNAL \cite{Kashikhinbs}, and the  DCC017 magnet built and tested by BNL \cite{Wandererqq}. The latter continues to be in use, serving as a magnet facility for a variety of high field ($\sim10$ T) tests for experiments for collaborators \cite{Wanderer4qz}.
    \item \say{Block Dipole} magnets; similar to the common-coil layout, but energized in \say{same polarity}, resulting in a single, high field bore. To maximize field while allowing access for the particle beam, the ends are \say{flared}. The concept was most thoroughly explored by LBNL in the mid 2000's, culminating in the \say{HD3} magnet that achieved $13.8$ T at $4.2$ K \cite{Prestemonr5g}. The technology was ultimately utilized in the CERN \say{FRESCA-II} magnet, which achieved $14.6$ T with a large, $\sim 100$ mm bore \cite{Willering8fg}. The magnet is not designed as an accelerator magnet, but rather serves as a test facility for superconducting cables.
\end{itemize}
We note that all of the above magnet concepts continue to be of interest, and in fact all are being pursued to some degree by institutions around the world. The lessons-learned from the research on these magnets formed the foundation for the research goals and magnet development roadmap for the US Magnet Development Program (MDP), which was created by DOE-OHEP in 2015 following the last Snowmass and P5 process, and motivated in particular by recommendations from the HEPAP Accelerator \randd Subpanel Report \cite{kzd}.

\begin{figure}[h!]
\centering
        \includegraphics[scale =0.7, angle=0]{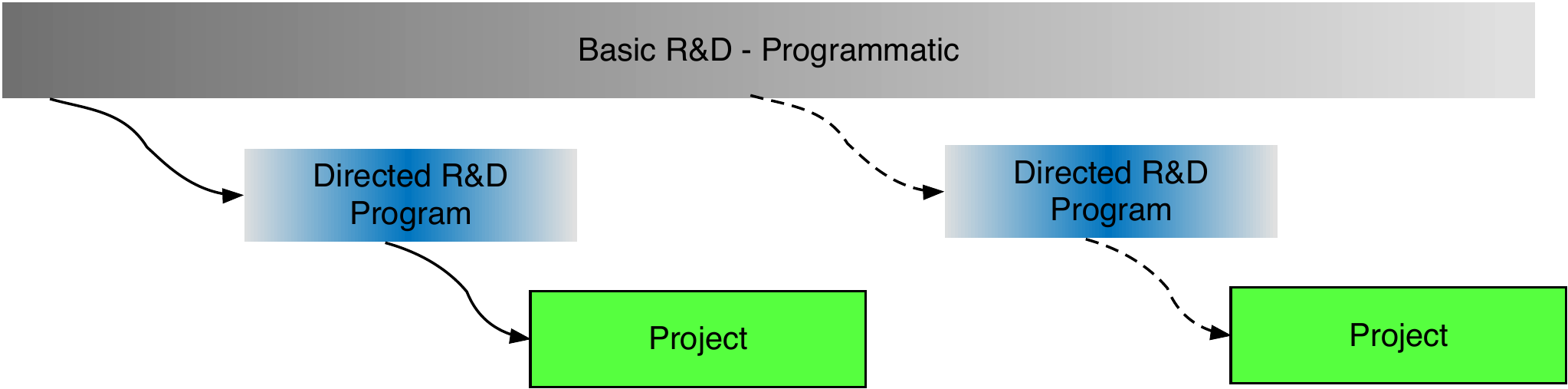}
       \caption{The DOE approach to advanced superconducting magnet technology has supported long-term \randd as well as, when appropriate, directed R$\&$D; an example of the latter was LARP, which, among other accelerator developments, prepared \nbsn quadrupole technology to enable the LHC luminosity upgrade (HiLumi) currently underway. DOE-OHEP is now contributing quadrupole magnets for HiLumi under the HL-LHC AUP project.}
       \label{fig:plot1}
\end{figure}

It is important to note that the US Magnet Development Program was created as a general \randd program, within the research arm of DOE-OHEP, and tasked with advancing accelerator magnet technology for future colliders, irrespective of  requirements of a specific collider. DOE-OHEP has an excellent track record of supporting long range R\&D, and initiating "Directed" \randd - essentially technology readiness programs - when it becomes apparent that a specific project, and hence project technology need, is on the horizon. The LHC Accelerator Research Program (LARP), and its follow-on DOE 413.3b project, HL-LHC AUP, serve as excellent examples (see Fig. \ref{fig:plot1}). LARP benefitted significantly from the expertise and experience from the magnet programs described above, which had performed critical exploratory research into the use of \nbsn for accelerator magnets, with many of the record dipole magnets described above having occurred prior to the onset of LARP. Of course the HL-LHC AUP project, currently underway, depended intimately on the developments from LARP, which largely developed and solidified the design and early demonstrations of the quadrupole magnets for the project.

Here we advocate for a strong, effective general \randd magnet program that develops high field magnet technology, guides the developments of advanced superconductors, and lays a foundation of scientific and engineering knowledge that provides lasting value for any future HEP collider - i.e. the US Magnet Development Program. A strong \randd program serves as the foundation needed for future directed \randd programs that target specific collider designs or that focus on cost reduction and scale-up. Examples of possible directed \randd programs are provided in complementary whitepapers \cite{apollinari1}, \cite{ambrosio1}.

\begin{tcolorbox}[colback=yellow!10!white,
                     colframe=black!75,
                     coltitle=black,
                     colbacktitle=black!10,
                     fonttitle=\bfseries,
                     title=Statement \thestatements : Complementarity of General and Directed R$\&$D,  
                     center, 
                     valign=top, 
                     halign=left,
                     before skip=0.8cm, 
                     after skip=1.2cm,
                     center title, 
                     width=15cm]
 
     The DOE-OHEP paradigm of long-range R$\&$D to advance magnet technology, coupled, when appropriate, with Directed-R$\&$D that demonstrates feasibility and prepares magnet technologies for projects, has proven extremely effective.
  \end{tcolorbox}
  \stepcounter{statements}

\section{The DOE-OHEP US Magnet Development Program}
\label{section:MDP}
The DOE-OHEP created the US Magnet Development Program in late 2015, pulling together experienced teams from DOE laboratories and the National High Magnetic Field Laboratory in a collaboration to advance  accelerator magnet technology for  HEP needs. As an element of the HEP General Accelerator \randd program, MDP is tasked with exploring the limits of high field accelerator magnet technology, without being constrained by the specific requirements of any given potential project.

The MDP is funding-limited and has by necessity - and appropriately - developed a focused mission and prioritized set of goals. Many excellent additional goals and research directions have been identified, but by necessity are not currently supported. Since its inception, the program has worked to streamline its operations and demonstrate an efficient and effective \randd operation, leveraging significantly synergistic activities, e.g. with Fusion Energy Sciences (FES), with the magnet \randd activities of the NHMFL supported by NSF, and with industry, in particular DOE's Small Business Innovation  Research (SBIR), as well as international collaborations. The program has oversight through a longstanding Technical Advisory Committee, composed of internationally recognized - and international - membership, as well as a Steering Council composed of laboratory leadership from the MDP collaborating institutions. 

\subsection{Scale of investment}
The MDP has made significant technical advances  and has demonstrated leadership in the field. Moving forward, based on our experience to-date, a general \randd program, i.e. MDP, would be most effective if funded at a level of $\sim \$15$M per year; to support the program's needs for superconductor, an additional $\sim 20$\%, i.e.  $\sim \$3$M, should be allocated for  procurement of industrial superconductor. Alternative mechanisms to obtain superconductor for programs such as MDP are explored in a companion whitepaper \cite{cooley1}. This investment is less than that proposed - and needed - by associated directed \randd programs, since directed \randd must address critical elements such as technology scale-up, value engineering, and reproducibility, all of which require more investment. We note that general \randd - MDP - invests heavily in resources - people - and hence serves as an incubator and repository of expertise for the community.

\subsection{Vision and goals aligned with HEP needs}
As a National Program composed of multiple DOE Laboratories and University members, the US Magnet Development Program (MDP) aspires to provide broad leadership in accelerator magnet technology \cite{Go16}. The vision of the MDP is to:
\begin{enumerate}
\item	Maintain and strengthen US Leadership in high-field accelerator magnet technology for future colliders;
\item	Further develop and integrate magnet research teams across the partner laboratories and US Universities for maximum value and effectiveness to MDP;
\item	Identify and nurture cross-cutting / synergistic activities with other programs (e.g. Fusion), to more rapidly advance progress towards our goals.
\end{enumerate}
These three core vision elements provide focus and direction to the program, while guiding interaction with other DOE-SC offices, international partners, and industry, to further the mission of the MDP. 
The overarching goals of the program remain unchanged after the program’s first four years:
\begin{itemize}
\item	Explore the performance limits of Nb$_3$Sn accelerator magnets, with a sharpened focus on minimizing the required operating margin and significantly reducing or eliminating training
\item	Develop and demonstrate an HTS accelerator magnet with a self-field of 5 T or greater, compatible with operation in a hybrid HTS/LTS magnet for fields beyond 16 T
\item	Investigate fundamental aspects of magnet design and technology that can lead to substantial performance improvements and magnet cost reduction
\item	Pursue Nb$_3$Sn and HTS conductor \randd with clear targets to increase performance, understand present performance limits, and reduce the cost of accelerator magnets
\end{itemize}

 The original 2015 MDP roadmaps were updated in 2020  \cite{Zlobinuy9}, to take into account progress that had been made by the program and to further strengthen and integrate the multi-lab collaboration. The major themes for the updated roadmaps include:
\begin{itemize}
\item	Explore the potential for stress-managed structures to enable high-field accelerator magnets, i.e. structures that mitigate degradation to strain-sensitive \nbsn and HTS superconductors in high-field environments; 
\item	Explore the potential for hybrid HTS/LTS magnets for cost-effective high field accelerator magnets that exceed the field strengths achievable with LTS materials;
\item	Advance magnet science through the rapid development and deployment of unique diagnostics and modeling tools to inform and accelerate magnet design improvements;
\item	Perform design studies on high field accelerator magnet concepts to inform DOE-OHEP on further promising avenues for magnet development;
\item	Advance superconductors through enhanced performance, improved production quality, and reduction in cost - all critical elements for future collider applications.
\end{itemize}
These themes are consistent with the original US MDP goals and leverage the major advances the US MDP has achieved to date in advancing superconductors, developing core HTS magnet technologies, and demonstrating record \nbsn accelerator magnet performance. Together, the themes form the foundation for a program that will maintain US leadership in developing advance accelerator magnet technology for the years to come.
\begin{tcolorbox}[colback=yellow!10!white,
                     colframe=black!75,
                     coltitle=black,
                     colbacktitle=black!10,
                     fonttitle=\bfseries,
                     title=Statement \thestatements : The US MDP is the result of a recommendation from the 2015 HEPAP Accelerator \randd Subpanel,  
                     center, 
                     valign=top, 
                     halign=left,
                     before skip=0.8cm, 
                     after skip=1.2cm,
                     center title, 
                     width=15cm]
 
     The US MDP is a long-term HEP GARD funded program that integrates leading DOE laboratory and University teams to advance accelerator magnet technology in support of HEP mission needs, addressing recommendations from the last Snowmass process and in particular the follow-on 2015 HEPAP Accelerator \randd Subpanel report.
  \end{tcolorbox}
  \stepcounter{statements}
  
\begin{tcolorbox}[colback=yellow!10!white,
                     colframe=black!75,
                     coltitle=black,
                     colbacktitle=black!10,
                     fonttitle=\bfseries,
                     title=Statement \thestatements : A general magnet \randd program is a cost-effective investment for DOE-OHEP,  
                     center, 
                     valign=top, 
                     halign=left,
                     before skip=0.8cm, 
                     after skip=1.2cm,
                     center title, 
                     width=15cm]
 
     General magnet R\&D, when well coordinated and managed and fully leveraging synergistic activities, is an excellent investment for high energy physics. The experience from MDP indicates that the scale of investment needed  is $\sim\$15M$ per year, with an additional $\sim \$20$\% (i.e. $\$3$M) in conductor procurement from industry to support the magnet needs. 
  \end{tcolorbox}
  \stepcounter{statements}
  
\subsection{Managing magnetic forces in high-field accelerator magnets}
To mitigate coil stress values at high fields and/or large apertures, stress-management (SM) concepts for various magnet coil geometries have been proposed \cite{Gaedkee4, Wang68, Zlobin1, Zlobin2}. The Canted-Cosine-Theta (CCT) dipole concept, under development at LBNL and PSI, is based on tilted solenoid coils. A Stress-Management Cosine-Theta (SMCT) dipole concept, in progress at FNAL, is built on the traditional cosine-theta magnet technology. The SMCT and CCT designs complement each other and address the question of whether stress-managed structures can fulfill their promise of breaking the traditional scaling of coil stress with field. If so, SM would enable high field magnet technology with stress/strain sensitive \nbsn. The same principle could be then applied to other stress/strain sensitive superconductors, such as High Temperature Superconductors (HTS). Together, the stress-managed magnet concepts are designed to address the key questions and provide capabilities – i.e. strong dipole fields and large bore – needed for HEP accelerator facilities. These concepts are also important for the development of very high-field hybrid dipole magnets with HTS coils as inserts and \nbsn coils as outserts. Integration of technical expertise and capabilities from the U.S. and EU laboratories, including areas such as magnet design, fabrication infrastructure, instrumentation, test facilities and test data analysis, will increase the efficiency and outcome of each magnet program.

\subsubsection{Approaches to stress-management}
Currently, as part of the US Magnet Development Program, two different stress-management approaches are being pursued. These are the Canted-Cosine-Theta (CCT) and Stress-Management Cosine-Theta (SMCT) approaches. These approaches share the common stress interception scheme of using \say{ribs} to transfer Lorentz forces to the \say{spar} portion of the mandrel. The SMCT method intercepts the force on blocks of conductor while the CCT method intercepts the force on every turn of the coil.  The SMCT winding geometry resembles that of a conventional cosine-theta magnet, while the CCT geometry consists of pairs of canted solenoids with opposing canting direction.  The key features of each design are further described in the following sections.

\begin{figure}[h!]
\centering
        \includegraphics[scale =0.8, angle=0]{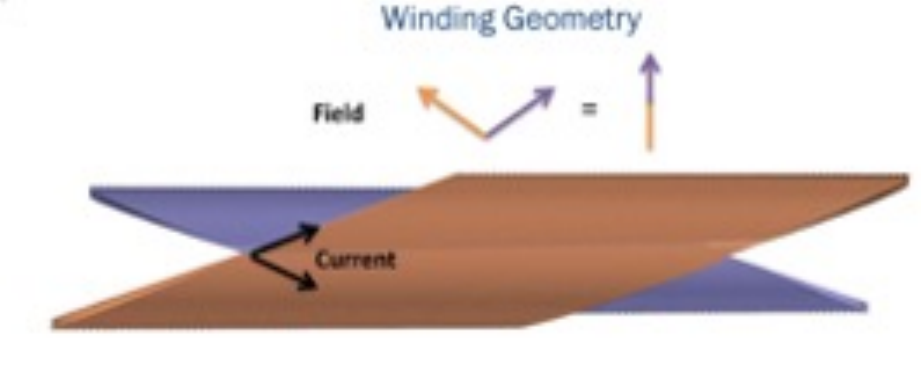}
          \includegraphics[scale =0.8, angle=0]{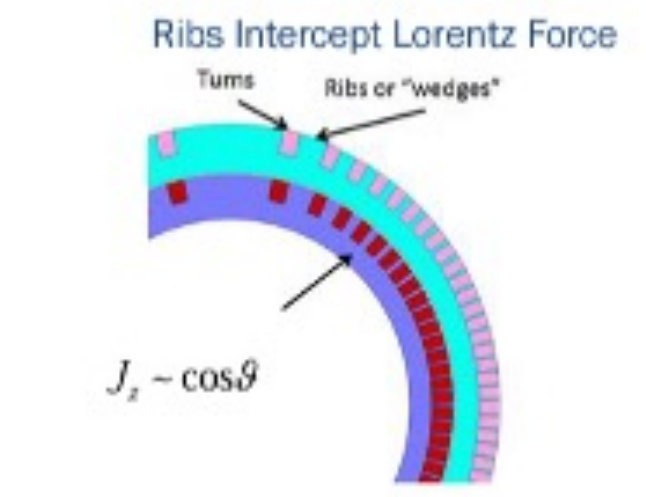}
       \caption{The "Canted Cosine Theta" (CCT) concept, wherein the Rutherford cable is directly supported in grooves machined into a support structure. The tilted winding results in a solenoidal field component, which is compensated by the next layer in the structure. The cross-section (right) shows the intrinsic ``Cos($\theta$)" nature of the resulting current distribution, resulting in excellent field quality.}
       \label{fig:CCT-concept}
\end{figure}

\paragraph{Canted Cosine Theta (CCT)}
Figure \ref{fig:CCT-concept} (left) shows the winding geometry for the CCT configuration. This consists of two tilted solenoids with opposing tilt orientation. The current direction is such that the dipole field component of each layer is additive, while the solenoidal field component is canceled. Figure \ref{fig:CCT-concept} (right) shows the cross-section of the coils. The intrinsic \say{Cos($\theta$)} nature of the conductor distribution leads to excellent field quality in the CCT dipoles. The internal support structure of the CCT dipole can also be seen in Figure \ref{fig:CCT-concept} (right). The Lorentz force at each turn is intercepted and partially transferred to the spar (the area in the mandrel below the ribs). This can lead to a significant reduction in the peak stress on the conductor. Another key advantage of CCT magnet technology is the ease of fabrication. A minimal number of parts are required, since many key features are incorporated into the mandrels. Since the mandrel also defines the basic geometry, this leads to greatly simplified reaction and impregnation tooling. The simplification of the manufacturing process could lead to significant cost reductions as the technology is further developed.\\

\begin{figure}[h!]
\centering
        \includegraphics[scale =0.8, angle=0]{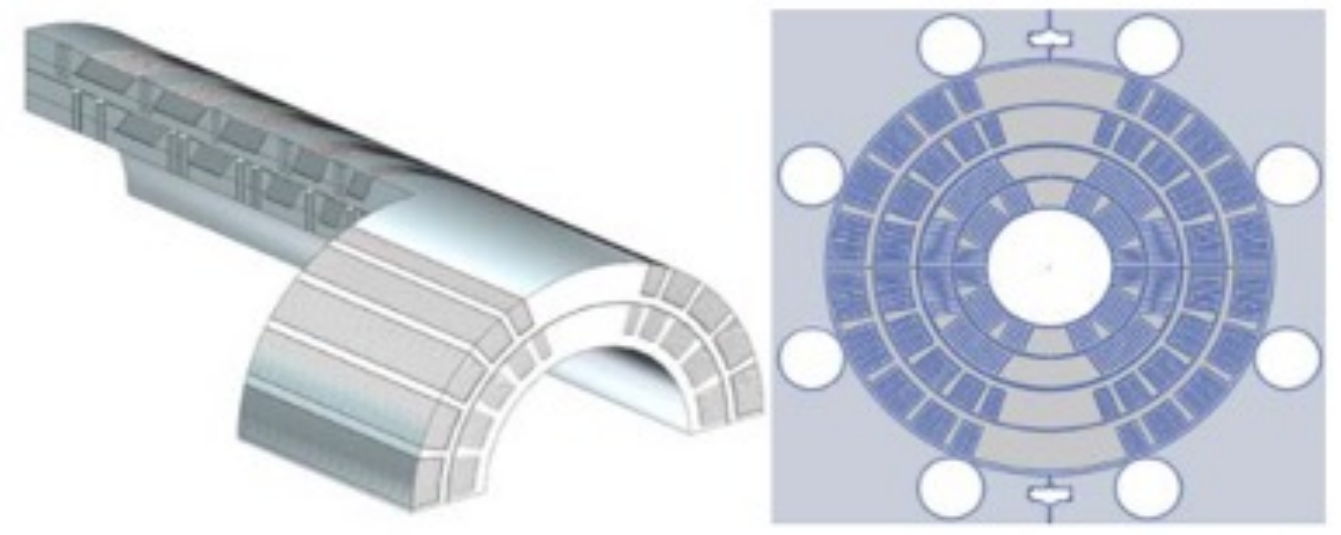}
       \caption{The Stress Managed Cosine Theta (SMCT) concept, wherein groups of  Rutherford cables are directly supported in grooves machined into a support structure. The groupings are designed such  that the cross-section (right) emulates a Cos($\theta$) current distribution, resulting in excellent field quality.}
       \label{fig:SMCT-concept}
\end{figure}

\paragraph{Stress-Managed Cosine Theta (SMCT)}
The winding geometry for SMCT magnets follows a traditional Cosine theta approach with conductor blocks that are separated by wedges. However, the necessary parts are incorporated into a solid mandrel internal structure as shown in Figure \ref{fig:SMCT-concept}. The mandrel contains the ribs / wedges and a spar to partially intercept the azimuthal Lorentz force on the coil blocks. As with the CCT concept, this can lead to significant reduction in the peak conductor stress. The two-dimensional nature of  the straight section in the SMCT leads to improved conductor efficiency when compared to the CCT approach. As is the case for CCT, the SMCT method consolidates the significant quantity of parts, common in more traditional magnet approaches, into one mandrel structure. This can lead to simplified winding, reaction, and impregnation processes and tooling. As with CCT, this could lead to significant cost savings as the technology becomes more mature.

\subsubsection{Areas of investigation for stress-management methods}
In order to reduce conductor stress, stress-management approaches introduce  internal structures and additional interfaces when compared to more traditional magnet designs. Furthermore, the mandrels which incorporate this structure can have fairly complex geometries. In order to take full advantage of these technologies, further understanding and development in several areas is necessary. Below is a list of some of the primary issues that are being investigated to develop these stress-management approaches:
\begin{itemize}
\item   Continue development of mandrel fabrication methods including exploration of conventional machining methods as well as additive manufacturing approaches;
\item   Investigate approaches for scale up to larger magnets including segmentation approaches;
\item	Explore new methods for layer-to-layer interface coupling, especially for the CCT approach where the coils have a closed structure;
\item	Investigate and optimize external structures for use in stress-managed magnets with increases internal structure rigidity. 
\item	Understand the influence of additional interfaces on stress-limits and training under different interface conditions;
\item	Understand the evolution of interface behavior, and the influence on training, due to interface damage in current and past magnet tests;
\item	Develop methods to improve control over interfaces for the desired behavior to reduce training and avoid unwanted stress re-distribution onto the conductor;
\item	Develop efficient and cost effective fabrication methods using minimal tooling to reduce overall magnet fabrication cost and complexity;
\item   Explore the limits of stress-management approaches on high field / large bore demonstrators.
\end{itemize}

\subsubsection{Development plans for stress-management methods}
The design of several magnet configurations using SMCT technology has been performed over the last few years \cite{Zlobin1,Zlobin2}. This includes the magnetic and mechanical design and analysis. Significant developments have also been made on fabrication technologies, specifically with regards to additive mandrel manufacturing \cite{novitski_using_2021} and winding methodologies. 

The CCT program has been ongoing since the start of the US MDP. The early focus has been on developing the methods and technology for CCT magnet design, analysis, fabrication, and testing. This early effort has culminated in the test of three two-layer \nbsn CCT dipoles with 90 mm aperture and a bore field of approximately 10 T at the short sample limit \cite{arbelaez_status_2022}. Each magnet in the series led to improved understanding and methods for CCT technology. The last magnet in the series reached approximately 88\% of the short sample limit. This effort has led to the next steps in CCT magnet development which are focused on demonstrating higher field / larger aperture operation and training understanding  / reduction. 

\paragraph{Small Scale Testing}
The additional interfaces that are introduced in stress-management approaches can be a source for energy release during magnet operation, which can lead to training quenches. Furthermore, if the interfaces evolve as the magnet is trained (i.e. by debonding and / or sliding of the conductor) this can lead to a redistribution of stress on the conductor. For example, a debonding event can lead to a reduction in shear stress and an increase in normal stress on the edge of the conductor. In order to probe / understand this behavior, small scale experiments that allow for systematic investigation of different parameters are desirable. Such experiments are currently ongoing with two different platforms. The BOX experiments being carried out at PSI and the University of Twente \cite{daly_box_2021} use cable in groove samples that are arranged in a serpentine pattern. The samples are then tested in a background field provided by an solenoid magnet. The focus of these experiments has been on probing training behavior with different impregnation materials and interface conditions. A complementary effort is going on at LBNL with subscale CCT magnets \cite{arbelaez_status_2022}. The subscale CCT magnets are designed to achieve  stress levels that ensure training at a similar load line margin when compared to the previously described two-layer \nbsn CCT series. This platform produces more comparable stress levels to larger CCT magnets than the BOX platform; however, this comes at the expense of higher complexity and slower turn around time. Similar to the BOX experiments, the aim of the subscale CCT program is to investigate how changes to structural conditions, impregnation materials, and interface conditions affect training and performance limits. Future testing at a subscale level could also include the SMCT geometry, in order to probe how training is affected when multiple turns lie inside of the mandrel grooves in a more traditional \say{Cos($\theta$)} geometry.

\paragraph{Large Scale Demonstrators}
Large scale demonstrators for stress-managed magnets are currently planned as part of the 2020 updated US MDP roadmap. For the SMCT approach, the roadmap R\&D goals are to demonstrate a bore field up to 11 T at 1.9 K with 120-mm aperture in two-layer \nbsn dipole magnets with stress-managed coils; and to demonstrate up to 17 T at 1.9 K with a 60-mm aperture in a four-layer \nbsn dipole magnet with stress-managed outer coils \cite{Zlobin1,Zlobin2,juchno_mechanical_2019}. For the CCT approach, the roadmap R\&D goals are to demonstrate bore field up to 13 T in a 120-mm aperture in a four-layer \nbsn CCT model magnet \cite{brouwer_design_2022}. Both of these efforts are closely aligned with the HTS programs as they can serve as the background field for HTS insert tests. This will allow for probing of HTS magnet behavior under high magnetic field and high stress states. The technology development of stress-management techniques for \nbsn is highly applicable to HTS conductors due to their similarly brittle nature. Stress-management methods are also currently being applied to HTS conductors \cite{Fajardo}, \cite{osti_1545078}, and will be critical for high field hybrid dipole magnets.

\begin{tcolorbox}[colback=yellow!10!white,
                     colframe=black!75,
                     coltitle=black,
                     colbacktitle=black!10,
                     fonttitle=\bfseries,
                     title=Statement \thestatements: Limiting magnetic force accumulation is essential to access high field magnets,
                     center,
                     valign=top, 
                     halign=left,
                     before skip=0.8cm, 
                     after skip=1.2cm,
                     center title, 
                     width=15cm]
 The concept of stress-management, i.e. intercepting magnetic forces to structural members before they can accumulate to damaging levels, promises a path towards very high field accelerator magnets.  
  \end{tcolorbox}
 \stepcounter{statements}
 
\subsection{The promise of high-temperature superconductors for HEP}

High Temperature Superconductors (HTS) have unique properties that are of particular interest to high energy physics applications. First, the ability to carry transport current at elevated temperature, above the traditional $1.8$ K or $4.2$ K associated with the liquid helium operation of low-temperature superconductors, opens the possibility of magnets operating at temperatures associated with other liquid cryogens, or in special cases with the use of cryogen-free cryostats cooled with cryocooler systems. The advantage of higher temperature operation may include higher overall efficiency (i.e. associated with Carnot efficiency), which can be particularly important in cases where heat loads on the magnet cold mass are significant. 

Second, and perhaps most importantly for collider applications, HTS materials have the additional property of a very high upper critical field $B_{c2}$; both Bi2212 and REBCO have $B_{c2}\sim100$ T, which opens the door to potentially very high field magnets. Attaining these high fields, however, requires surmounting significant challenges - learning how to work with the materials, understanding their behavior under stress and strain,  managing the tremendous forces that are created at high magnetic field, and developing methods to protect the magnets from damage if and when some part of the superconductor loses its superconducting state (i.e. a "quench"). Addressing these challenges is a central theme of the US MDP. 

While active material research continues increasing the potential of \nbsn conductors  towards  generating a  dipole  field  of  $20$  T  at  $4.2$  K  \cite{Xu14}, \cite{Ba19}, \cite{Sa15}, high-temperature superconductors (HTS) \cite{Ma12} are required to generate dipole fields at and beyond $20$ T. Two main candidate HTS conductors are Bi-2212 round wires \cite{La14} and REBCO coated conductors \cite{Se12}. 

\subsubsection{Bi2212}
The superconductor Bi2212 is manufactured using a powder-in-tube technique, a well-established superconductor architecture that allows for a multi-filament composite wire to be manufactured \cite{La14}. This has a number of advantages:
\begin{itemize}
    \item Because it is multifilamentary and isotropic, magnetization losses are controlled and the conductor is "well-behaved" for accelerator applications where field quality is a concern.
    \item because it is a round wire, the well established HEP route to scale up current, through the use of Rutherford cables, can be applied. Rutherford cables enable efficient use of superconductors while minimizing the number of turns in a magnet, thereby reducing the magnet inductance; the inductance is critical parameter in magnet protection, i.e. the ability to extract the stored magnetic energy in the case of a quench.
    \item Because it builds on established wire drawing fabrication processes, there is potential for some degree of scale-up in the production capacity, and industry has tooling and procedures that can be readily applied to Bi2212 wire production.
\end{itemize}
Furthermore, the transport current of Bi2212 has seen significant improvement over the last decade, and the wire is now competitive with \nbsn above $\sim 14$ T, making it a very attractive candidate for accelerator magnets beyond $\sim 16$ T.   

Despite these advantages, the conductor currently is being actively developed almost exclusively by HEP and by the high field NMR community. This is primarily due to complexities in its implementation in magnets - Bi2212 requires a high temperature heat treatment ($\sim 890$ $\degree C$) in a 1-bar Oxygen environment. Furthermore, to access the highest performance, the heat treatment must be performed at \say{overpressure}, e.g. at $\sim 50$ bar (still at 1-bar $O_2$). Finally, like all advanced superconductors, once formed the superconductor is strain sensitive and brittle, so handling of the superconductor must be minimized; for this reason magnet research leans  towards a "wind-and-react" approach. Evaluating the transport current strain sensitivity and  degradation limits, in particular under transverse pressure, is important for accelerator application.  The combination of issues described above puts constraints on the tooling design and material.

We are steadily progressing on addressing the technical challenges associated with the use of Bi2212 superconductor for accelerator magnets, including issues related to the reaction process such as insulations and conductor treatments that avoid Bi2212 leakage during heat-treatment, and magnet and tooling materials that are compatible with the reaction temperature and environment, and are now working towards first implementation of Bi2212 in a hybrid HTS/LTS magnet, where Bi2212 is applied in the inner coils that are subjected to high field, and the outer coils utilize Nb$_3$Sn, due to their higher efficiency at lower field.

A  review of the status of Bi2212 accelerator magnet research, and how the material can play an essential role for HEP in the coming decades, is provided in a companion whitepaper \cite{tengming1}.
\begin{tcolorbox}[colback=yellow!10!white,
                     colframe=black!75,
                     coltitle=black,
                     colbacktitle=black!10,
                     fonttitle=\bfseries,
                     title=Statement \thestatements: Bi2212 is a leading HTS material candidate for accelerator magnets that benefits from many of the conductor characteristics proven effective by HEP over the last decades, 
                     center,
                     valign=top, 
                     halign=left,
                     before skip=0.8cm, 
                     after skip=1.2cm,
                     center title, 
                     width=15cm]
 Key characteristics, including the fact it is produced as a multifilamentary round wire that can be readily made into a Rutherford cable, make Bi2212 a leading HTS candidate material for HEP applications. The challenge now is to translate the impressive progress in transport current achieved from commercial wires over the last decade into accelerator magnet performance. 
  \end{tcolorbox}
 \stepcounter{statements}

\subsubsection{REBCO}
Of the two HTS candidate conductors, REBCO is the greater departure from LTS cconductors because coated conductors are available only in tape form, with a high aspect ratio of at least 10. The brittle REBCO layer is already in the conductor that is delivered from vendor to magnet builders. Handling of the tape for magnet fabrication and operation must be performed carefully so as not to exceed a tensile strain of only about 0.6\%. There are several reasons why we painstakingly learn how to use this material:
\begin{itemize}
    \item Potential low cost due to the low raw material cost, even if today's cost remains very high;
    \item Capability to generate high magnetic fields over a broad temperature range of 1.9 – 40 K. 
\end{itemize}

\begin{figure}[h!]
\centering
        \includegraphics[scale =0.9, angle=0]{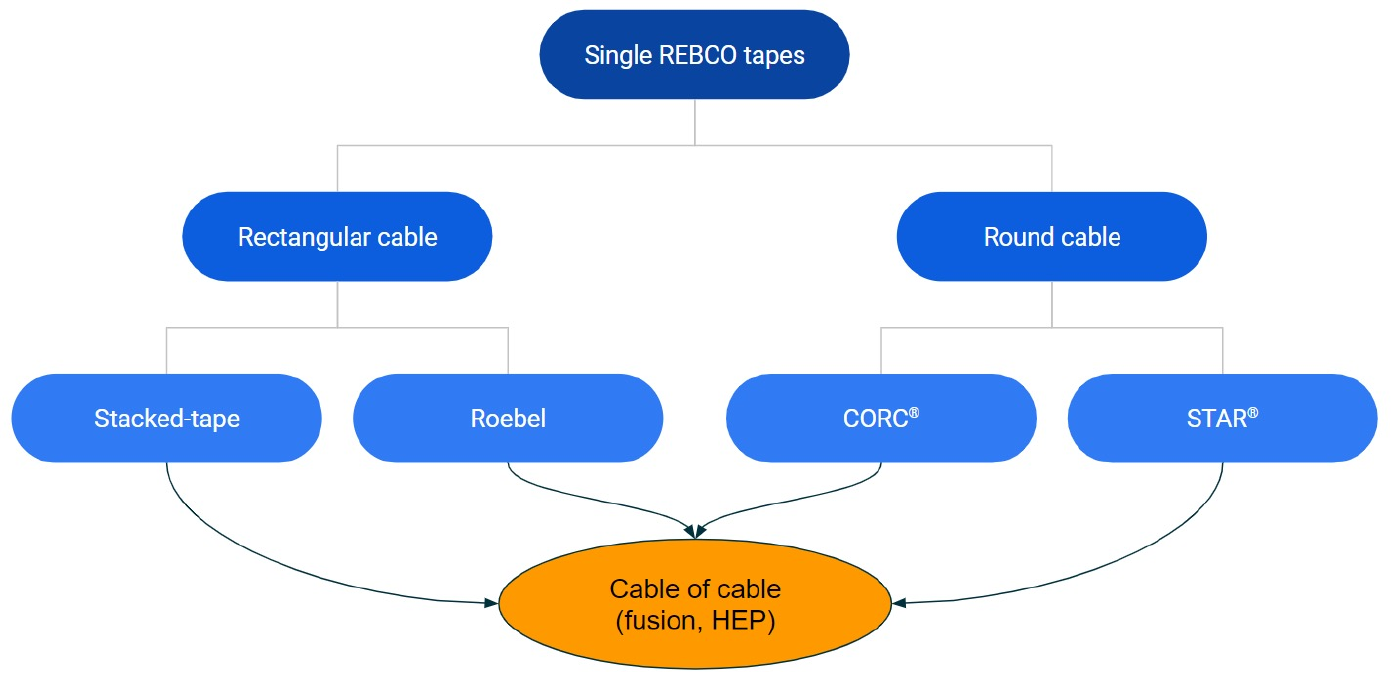}
       \caption{REBCO technology has a unique challenge: we need to build multi-tape conductor from a highly-aspected tape. Several concepts exist for a multi-tape REBCO cable that can be used for high-field dipole magnets. A detailed comparison regarding the merits and drawbacks of each concept is not yet available, partially due to limited magnet results, and a down-select is not yet feasible. The U.S. MDP is focused on the round-wire architecture while the European counterpart has been traditionally focused on the rectangular cable architecture based on stack of tapes. }
       \label{fig:Wang1}
\end{figure}

These two factors could have a profound impact on future collider projects that require high-field magnets. The recent European Accelerator \randd Roadmap has well explained the rationale for their focus on REBCO \cite{EU22}. In a companion white paper submitted to the Snowmass Community Study, we will explore these topics in some details and offer additional opinions \cite{wangrebco}.
The U.S. MDP has long recognized the potential of REBCO. One of the overarching goals of MDP is to demonstrate a dipole field of $5$ T or greater, and measure its field quality \cite{Go16}. The recently updated MDP roadmap has pointed to the hybrid LTS/HTS dipole magnet as a vehicle to further study, develop and exploit the potential of HTS insert magnet technology \cite{So20}. To reach these goals, we plan the following strategy for REBCO magnet technology development, recognizing that REBCO is the least familiar one among all the existing technical conductors for HEP applications: 
\begin{itemize}
\item	Strongly couple the development of REBCO conductor and magnet technology. Use the magnet results as a critical feedback to the conductor vendor that, in turn, will help improve the magnet performance. Conductor architecture that is most suitable for high-field dipole magnets is a question to be addressed (see Fig. \ref{fig:Wang1}). We currently focus on the round-wire cable architectures that are pursued by  U.S. technology startups, but keep an eye on alternative conductor architectures that are pursued by our European colleagues. The round wire architecture is essentially isotropic in terms of mechanics and magnetics, making it more readily scalable and transportable to new magnet architectures. 
\item	Take incremental but, hopefully, fast steps to grow the magnet technology. Stick to the concept of \say{minimum viable magnet} with a focus on generating higher dipole fields. Understanding implications but avoiding premature optimization on important issues such as field quality and quench protection. Building upon the experience and lessons learned from the previous steps, introduce and explore a limited number of  new features in the next magnets towards a full set of magnet technology that can yield the ultimate dipole fields. 
\item	Leverage MDP/CPRD, DOE SBIR and synergetic fusion programs to collaborate with potential vendors to advance their product towards magnet conductors and to develop relevant and critical magnet technologies that can help MDP succeed in its mission. 
\end{itemize}

\begin{figure}[h!]
\centering
        \includegraphics[scale =0.9, angle=0]{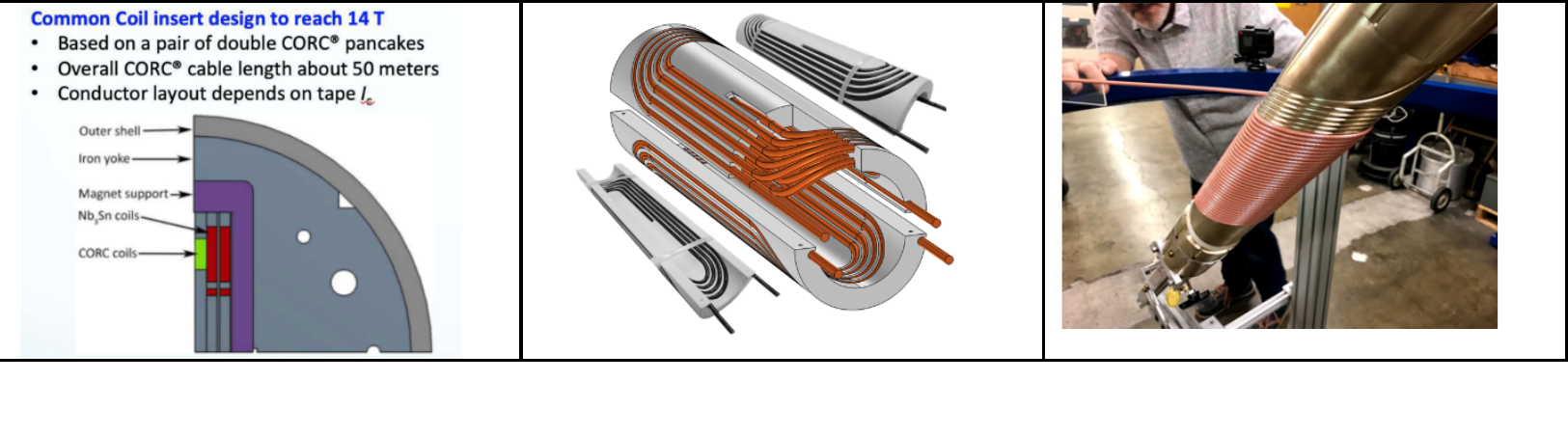}
       \caption{Three magnet configurations that are currently being pursued by collaborating laboratories at MDP. Left: common coil. Middle: COMB. Right: CCT. All three configurations feature stress management and are developed for round REBCO wires. }
       \label{fig:Wang2}
\end{figure}
   
   In the past few years, the REBCO community has  investigated different dipole magnet configurations using CORC$^{\circledR}$ wires and progressed towards a high-field dipole magnet technology. The three magnet configurations are common coil being developed at BNL \cite{CC}, Conductor on Molded Barrels (COMB) at FNAL \cite{COMB}, and CCT at LBNL \cite{C0}, \cite{C1}, \cite{C2}. Figure \ref{fig:Wang2} shows the three concepts. All magnet configurations feature stress-management to better enable high-field applications. Subscale models of both common coil and COMB concepts are being developed with magnets of a larger scale planned for development and testing within the next 12 months. The latest CCT magnet generated a peak dipole field of 2.9 T at 4.2 K in  \cite{C2}, with ongoing work to develop a magnet to 5 T at 4.2 K. 
   
   MDP is currently collaborating with the Colorado-based Advanced Conductor Technologies, LLC, the vendor of CORC$^{\circledR}$ wires to develop the various dipole magnet configurations. MDP is also evaluating the potential of STAR$^{\circledR}$ conductor from the Texas-based AMPeers LLC as an alternative round-wire architecture for REBCO tapes. 
In the coming years, the REBCO teams will continue focusing on increasing the dipole field that a REBCO magnet can generate as the single most important goal for two reasons. First, a higher dipole field is critical to support the strategic hybrid LTS/HTS dipole magnet configuration towards a dipole field of 20 T. The ongoing design analysis of the 20 T hybrid dipole magnet shows that to generate a 5 T dipole field from an HTS insert in a background field of 15 T, the insert needs to generate at least 10 T dipole field when being tested stand-alone \cite{Pa22}. This is nontrivial, considering today we are capable of generating no more than 5 T.  

Second, a REBCO magnet capable of higher dipole fields can emerge as a strong candidate for stand-alone applications. As mentioned earlier, REBCO magnets can in theory generate strong field ($>10$ T) at elevated temperatures. This has two implications.  First, it opens the possibility to operate without liquid helium, e.g. with cryocoolers. Second, a higher efficiency to extract heat compared to operation at 1.9 or 4.2 K. This can be very appealing for applications with high heat load, such as the proposed pp machine and muon colliders \cite{EU22}. 
We also note that REBCO is a focus of the rising fusion technology development as one of the enabling technologies for fusion \cite{Fusion}. REBCO is at the heart of the compact high-field fusion magnets that are being actively developed by several fusion technology companies that have attracted unprecedented public attention and private investment. A strong synergy between HEP and Fusion magnets naturally appears. The fusion application, if successful, can establish a sustainable market for REBCO conductors, which can help reduce today’s high conductor cost. The MDP recognizes the strong synergy and seeks to collaborate with our fusion colleagues in both public and private sectors to address the technology needs for high-field REBCO magnets. An excellent example is the development of a test facility dipole magnet that is jointly supported by DOE OFES and HEP. The magnet for the facility will be a state-of-art \nbsn block-design high-field dipole magnet that will provide a 15 T dipole background field for HTS insert and fusion cable testing. 

\begin{tcolorbox}[colback=yellow!10!white,
                     colframe=black!75,
                     coltitle=black,
                     colbacktitle=black!10,
                     fonttitle=\bfseries,
                     title=Statement \thestatements: REBCO has strong potential for HEP applications, 
                     center,
                     valign=top, 
                     halign=left,
                     before skip=0.8cm, 
                     after skip=1.2cm,
                     center title, 
                     width=15cm]
 Continuation and increase of the strategic investment in REBCO conductor and magnet technology development will be critical to establish  U.S. leadership in this future cost-effective high-field magnet technology that can benefit both HEP  and fusion energy applications. 
  \end{tcolorbox}
 \stepcounter{statements}

\subsection{Considerations for accelerator magnets beyond 20 T}
In superconducting dipole magnets, bore fields at the $15$ to $16$ T level are considered as the practical limit for \nbsn superconductor \cite{Sch1}. To further push the magnetic field of  dipole magnets beyond the \nbsn limits, High Temperature Superconductors (HTS) need to be considered in the magnet design. For accelerator magnets, the most promising HTS materials currently under consideration are Bi2212 \cite{La14} and REBCO \cite{Selva1}. However, their outstanding performance still comes with a significantly higher cost than Nb$_3$Sn. Therefore, an economically viable option of 20 T dipole magnets has to involve a \say{hybrid} approach, where HTS materials are used in the high field part of the coil with so-called \say{insert coils}, and \nbsn and Nb-Ti superconductors are adopted in the lower field region with so-called \say{outsert coils}. Preliminary design studies of 20 T hybrid dipole magnets were carried out in 2005 \cite{Mcintyre1} and in 2014-2016 \cite{Rossi1, ezio1, 10.1109/TASC.2015.2511927}, whereas a full HTS option was analyzed in 2018 \cite{Nugterenln4}. In 2015, a 20 T hybrid block-type design was presented by G. Sabbi, et al. in \cite{Wangneq}. A hybrid magnet was recently attempted by inserting a REBCO coil inside the FRESCA2 dipole magnet \cite{Canale}. Finally, REBCO inserts based on Roebel cables were fabricated and tested as part of the EUCARD2 Collaboration \cite{Nugteren}, \cite{Rossi2t4}. 

Here we describe a preliminary conceptual magnetic design analysis of a $20$ T hybrid dipole magnet for particle accelerators implementing \nbsn and HTS coils (Nb-Ti is not considered in this study). First, a description of the superconducting material properties and the criteria used in the different designs is provided. Then, the magnetic analysis of 3 types of coil lay-outs, 1) cos-theta, with and without stress management, 2) block-type, with and without stress management, and 3) common-coil, are presented. 

\begin{figure}[!ht]
\centering
        \includegraphics[scale =0.5, angle=0]{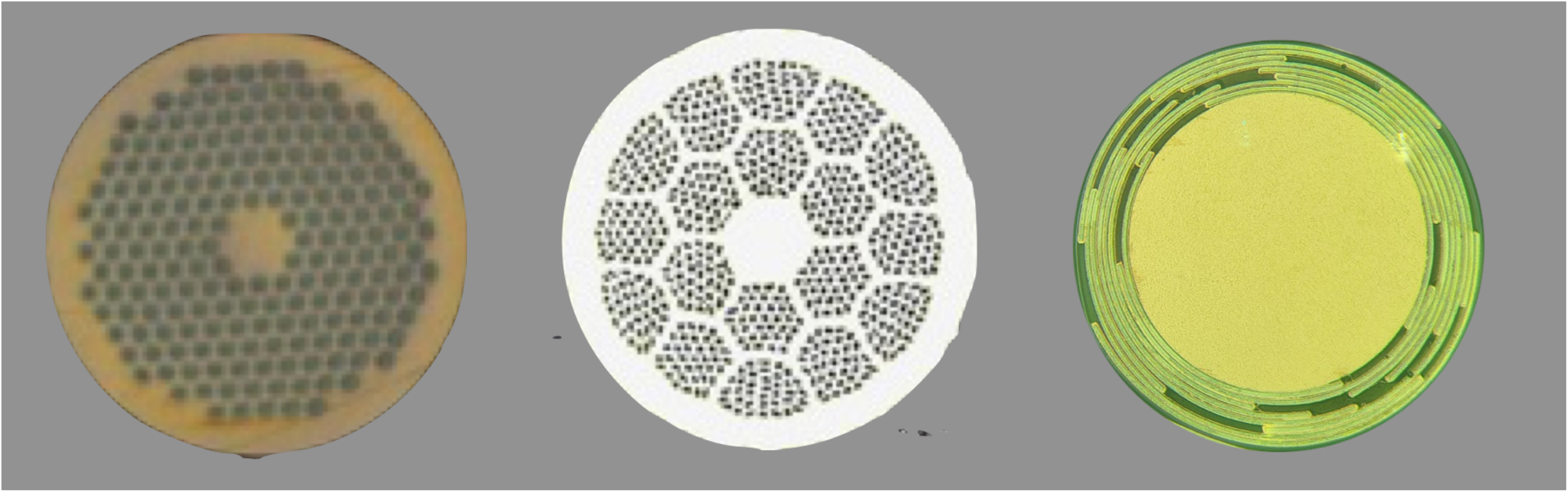},
       \caption{Cross-sections, not in scale, of Nb$_3$Sn (left, $0.85$ mm \diameter), and Bi2212 (center, $0.8$ mm \diameter) composite wires produced by Bruker-OST, and a REBCO CORC$^{\circledR}$ wire (right, $3.4$ mm \diameter) by ACT LLC.}
       \label{fig:superconductors}
\end{figure}

An example of the superconducting material properties considered for this preliminary analysis is shown in Fig. \ref{fig:superconductors}, with a Nb$_3$Sn Rod restack Process (RRP) and a Bi2212 strand produced by Bruker-OST, and a REBCO CORC\texttrademark{} wire produced by ACT LLC. For each material, the engineering current density $J_e$ used in the computations, where $J_e$ is the critical strand current divided by the strand cross-section area, is given in Table \ref{Tab:superconductor-props}.

 \begin{table}[htb]
	\caption{Superconductor Strand and Cable Properties Assumed}
	\label{Tab:superconductor-props} 
	\centering
	\begin{tabular}{lcccc}	
	\hline 
	 Parameter &	Unit	& Nb$_3$Sn&	Bi2212 &	REBCO  \\ 
	    \midrule
Strand diameter &	mm	& $0.7-1.1$	& $0.7-1.1$ &	$1.2-4.0$ \\
Cable width	& mm	& $7.8-26.2$	& $7.8-26.2$	& NA\\
Cable thickness	mm	& $1.2-2.0$	& $1.2-2.0$	& NA\\
$J_e$ (at $1.9$ K, $16$ T)	& A/mm$^2$	& 870	& 800	& 700\\
$J_e$ (at 1.9 K, 20 T)	& A/mm$^2$	& 360	& 740 &	590\\
$J_0$ / $J_e$	& 	$0.67$	& $0.67$	& $0.54$\\
       \bottomrule \hline
	\end{tabular}
	\end{table}

\subsubsection{Design Criteria}
The criteria defined in the conceptual design are summarized in Table \ref{Tab:design-criteria-20T}. The magnet must be able to produce 20 T in a 50 mm clear aperture with at least 15\% of load-line margin. This means that the \say{short-sample} bore field, i.e. the bore field achieved when the magnet reaches the current limits established by the conductor critical surface, is at least 23.5 T. The design must have all the geometrical harmonics field below 5 units at the nominal field and at 2/3 of the aperture radius (magnetization effects are not included at this stage). All the coils are assumed to be powered in series (a condition which impacts both the magnetic design and the quench protection system), and the hot spot temperature at quench is assumed to be limited to 350 K, both in the \nbsn \cite{ambrosiomax} and in the HTS coils \cite{Schwartz3wn}, \cite{Ishiyamatl5}, \cite{wangcrit}. In terms of stress limits, for the \nbsn we chose 150 and 180 MPa at 293 K and 1.9 K respectively, consistently with results published in \cite{10.1088/1361-6668/aab5fa}, whereas for the HTS we defined a preliminary and more conservative value of 120 MPa \cite{Dietdericht8}, \cite{Laank65}.

 \begin{table}[htb]
	\caption{Magnet design parameters used in the study}
	\label{Tab:design-criteria-20T} 
	\centering
	\begin{tabular}{lcc}	
	\hline 
	    \toprule
	   Parameter & Unit & Value \\ 	
	    \midrule
	   Operational temperature &	K	& $1.9$ \\
Operational bore field $B_{0_op}$	& T &	$20$ \\
Load-line margin	& \% & 	$15$ \\
Geometrical harmonics ($20$ T, $R_{ref}=17$ mm) &	unit	& $<5$ \\
Maximum Nb$_3$Sn coil eq. stress ($293$ K) &	MPa &	$150$ \\ 
Maximum Nb$_3$Sn coil eq. stress ($1.9$ K) &	MPa	& $180$ \\ 
Maximum HTS coil eq. stress ($293$ K, $1.9$ K) &	MPa	& 120 \\ 
Maximum hot spot temperature  &	K	& $350$ \\
       \bottomrule \hline
	\end{tabular}
	\end{table}

\subsubsection{Magnetic Design}
\begin{figure}[h!]
\centering
        \includegraphics[scale =0.52, angle=0]{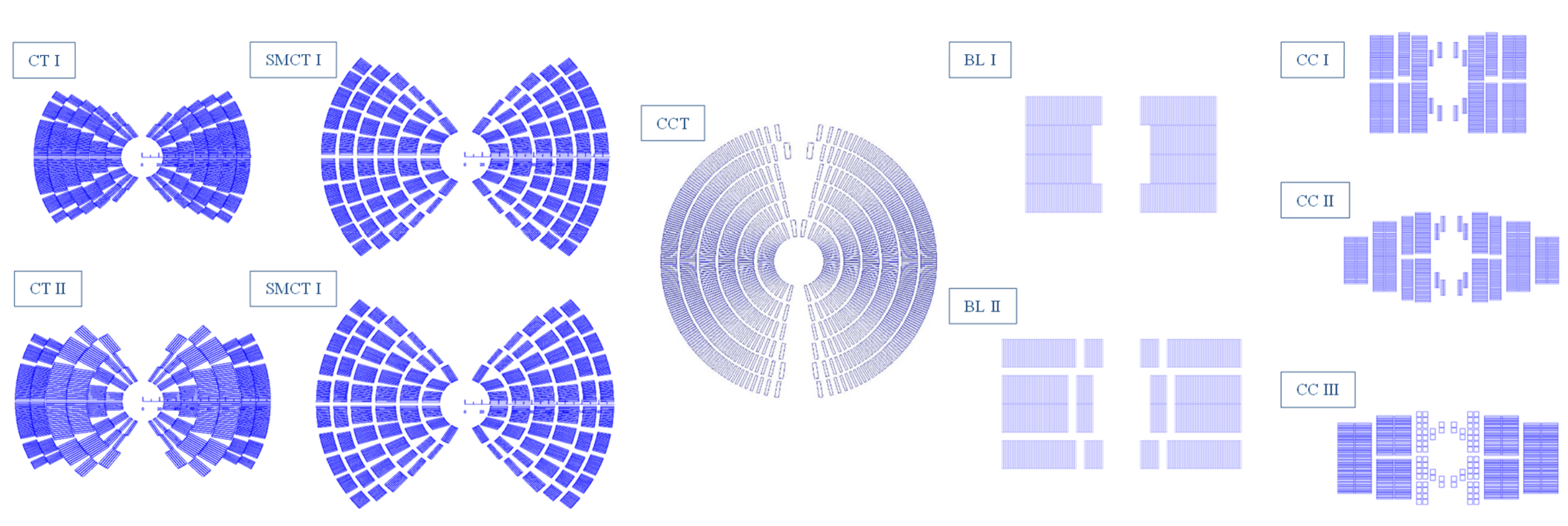}
       \caption{Cross-sections of 20 T hybrid dipole coils. The designs use consistent conductor properties, but are at different stages of the analysis in terms of field quality, mechanics, and quench protection: therefore, they are not comparable among them.  From left two right: Cos-theta (CT) design, with 4 (top) and 2 (bottom) layer Bi2212 coils; Stress management Cos-theta (SMCT) design, with 4 (top) and 2 (bottom) layers Bi2212 coils; Canted Cos-theta (CCT) design, with 4-layer Bi2212 coil; Block (BL) design, with and without stress management; Common Coil (CC) design, with Bi2212 (top, with 3 external \nbsn layers, and center, with 5 external \nbsn layers) and REBCO CORC$^{\circledR}$ coils (bottom, with 4 external \nbsn layers). For all the CC designs, only one aperture is shown.}
       \label{fig:20T-cross-sections}
\end{figure}

For a preliminary magnetic analysis of the 20 T hybrid magnet, we investigated different coil design options, all shown in Fig. \ref{fig:20T-cross-sections}. In particular, the following lay-outs were considered: traditional Cos-theta (CT) design, Stress Management Cos-theta(SMCT) design, Canted Cos-theta (CCT) design, Block (BL) design, and Common-Coil (CC) design. The main parameters are given in Table 3. The cables considered range from 9.1 to 21.6 mm width and from 1.5 to 2.1 mm thickness (including insulation). All the designs implement HTS Bi2212 Rutherford cables; REBCO CORC$^{\circledR}$ wire, with an assumed 7.5 x 7.5 mm$^2$ of dimension with insulation, was considered only for the common-coil, given the large bending radius of the latter. With a 20 T bore field, the operational current varies from 10.7 to 17.8 kA, and the peak field in the HTS and LTS coils is respectively 20.2-21.0 T and 12.7-17.0 T. The target load-line margin is achieved in the CT, SMCT, and BL designs, while in the others the margin is a few percentage points below target. In terms of field quality, design criteria are met by all the designs except the BL, which features geometric harmonics up to the 10 units level.  Finally, it is important to point out that only a preliminary investigation of the accumulated electro-magnetic (e.m.) forces in some of the designs was carried out, and a complete mechanical analysis aimed at bringing the stress in the HTS and LTS below the limits fixed in Table II has not been performed yet. Similarly, a full field quality and quench protection analysis has not been performed. Therefore, the designs depicted in Fig. \ref{fig:20T-cross-sections} represent only a first iteration and a starting point of the design, and, since they meet only part of the criteria, they are not yet comparable. 

\begingroup
\setlength{\tabcolsep}{1.5pt}
\begin{table}[htb]
	\caption{20 T Hybrid Magnet Parameters. \\ Definitions: "w"=width, "t"=thickness, "op"=operational, "ss"=short-sample.}
	\label{Tab:magnet-results} 
	\centering
	\scriptsize
\begin{tabularx}{\textwidth}{llllllllllll}
\hline \\
Parameter &	Unit	& CT I	& CT II	& SMCT I	& SMCT II &	CCT &	BL I	& BL II & 	CC I	& CC II	& CC III \\ \hline
Cable I w/t	&mm &	18.7/1.5	& 20.9/1.7	& 21.4/1.5	& 21.4/1.5	& 18.7/1.9	& 17.1/2.1	& 17.1/2.1	& 18.7/1.8	& 18.7/1.8	& 7.5/7.5 \\
Cable II w/t &	mm	&16.4/1.5	& 24.7/2.1	 &8.7/1.5	& 24.2/1.5	& -&	17.1/2.1	& 17.1/2.1&	13.6/1.9&	13.6/1.9&	21.6/1.9\\
Cable III w/t &	mm	&16.3/1.5	&18.0/1.5	&15.8/1.5&	15.8/1.5	&-&	-&	-&	-&	-&	-\\
Current$_{\textit{op}}$ &	kA&	10.7	&13&	11.4&	11.8&	12.8&	12.6&	12.2&	14	& 13.9&	17.8\\
B$_\textit{bore}^\textit{op}$	&T	&20&	20&	20.1	&20	&20&	20&	20&	20&	20&	20\\
B$_\textit{peak}^\textit{op}$ HTS/LTS	&T	&20.5/12.7&	20.3/16.1&	20.6/13.6&	20.6/16.0	&20.2/13.2&	20.6/15.1& 	20.9/15.2&	20.4/13.8&	20.2/13.7	&21.0/17.0\\
B$_\textit{bore}^\textit{ss}$ &	T&	24.4	&23.5	&24.4&	23.2	&23.4&	23.6&	23.6	&22.9&	23&	21.7\\
B$_\textit{peak}^\textit{ss}$ HTS/LTS&	T&	24.9/15.4&	23.8/17.7&	24.9/16.4&	23.8/18.4&	23.6/12.9&	24.3/17.7&	24.7/18.0&	23.3/15.7&	23.3/15.7&	24.7/18.2\\
Margin&	\%&	18/25&	21/15&	22/18&	20/15&	14/14&	21/17&	22/17&	13/13&	13/13&	15/7\\
Area cable HTS&	mm$^2$&	3241&	1494&	2091&	1527&	4490&	1360&	1500&	1290&	1154&	1012\\
Area cable LTS&	mm$^2$ &	2150&	6106&	3780&	5148&	4915&	4740&	4640&	2326&	2558&	4191\\
Coil width	&mm&	105&	129&	144&	149&	135&	80&	112&	70&	104&	106\\
Coil inner rad	&mm	&25&	25&	30	&30	&30&	35&	35&	25&	25&	25\\
       \bottomrule \hline \\
	\end{tabularx}
	\end{table}  
\endgroup
\paragraph{Cos-theta (CT), Stress Management Cos-theta (SMCT), and Canted Cos-theta (CCT) Designs}
For the CT options we considered a design with double layers coils, each wound with the same cable. This design choice avoids interlayer splices and has been implemented in most of the \nbsn CT coils fabricated so far (the only exception being the CERN-ELIN and UT-CERN dipole magnets \cite{Sch1}). The first design (top CT in Fig. \ref{fig:20T-cross-sections}) has 4 Bi2212 layers and 2 \nbsn layers, with a peak field in the \nbsn of 12.7 T and a total coil width of 105 mm. In a second design (bottom CT in Fig. \ref{fig:20T-cross-sections}) we reduced the Bi2212 layers from 4 to 2 by increasing the width of the \nbsn coils. As a result, the peak field in the \nbsn rose to 16.1 T and the total coil width rose to 129 mm. As shown in sector coils, a traditional CT design magnet aiming at 20 T is characterized by high coil stress in the azimuthal and radial directions. A possible alternative solution is the SMCT, where each layer is separated by 5 mm thick spars (or mandrels) and each cable block is separated by ribs, connected to the mandrel \cite{Zlobin2}. The implementation of stress intercepting elements results in an overall increase of coil width from 102-129 mm in the CT to 144-149 mm, and in the conductor area. A further step towards the reduction of the stress is done with the CCT design, where each turn is separated by spars and ribs [29]-[31]. The field quality is naturally achieved by superimposing the two tilted solenoids (see Fig. \ref{fig:20T-cross-sections} center). For the 20 T hybrid we chose a simple design with 4 Bi2212 layers and 2 \nbsn layers, all wound with a MQXF cable. The total area of the insulated cable (taken from a simple cross-section of the 3D design) is, as expected, larger than in the previous CT and SMCT designs. However, since the layer-to-layer splices are located in the coil ends, a full graded coil, with cables progressively smaller from the inserts to the outserts, can reduce significantly the coil size, and it will be the goal of the next step in the optimization.  

\paragraph{Block (BL) Design}
The Block design \cite{Wangqq}, \cite{8607102}, \cite{Lorin}  allows for a very efficient subdivision between the HTS and LTS coils, since the cables are aligned with the flux lines. Therefore, the area of Bi2212 in the block design shown in Fig \ref{fig:20T-cross-sections} (BL top design), is smaller than for the CT, SMCT and CCT options (see Table \ref{Tab:magnet-results}). Also, in terms of total conductor area the design is very compacted, despite the inclusion of a 10 mm thick internal support in the inner coil that increases the coil aperture to 70 mm (similarly to the coil design of FRESCA2 \cite{6423253} and TFD \cite{6423253} magnets).  However, as shown in \cite{10.1109/TASC.2022.3158643}, the peak stress in the coil, in particular because of the horizontal e.m. forces, can be as high as 280 MPa in the \nbsn and 160 MPa in the Bi2212 at 20 T. Therefore, an alternative has been considered where vertical and horizontal plates are included to intercept part of the e.m. forces.  In the bottom BL design in Fig. \ref{fig:20T-cross-sections}, intercepting plates separates Bi2212 and \nbsn coils: consequently, the coil in-creases in size, but the stress in the \nbsn and in the Bi2212 coils decreases to about 160 MPa and 140 MPa respectively at 20 T. 

\paragraph{Common-coil design}
On Fig. \ref{fig:20T-cross-sections}, 3 different common coil designs for the 20 T hybrid are shown. The common-coil design \cite{10.1109/TASC.2016.2636138}, \cite{10.1109/TASC.2018.2797909}, \cite{10.1109/TASC.2018.2813518}  is based on racetrack coils that, with a large bending radius in the ends, cover both magnet apertures (in Fig. \ref{fig:20T-cross-sections} only one aperture is shown). The large bending radius opens the possibility not only of implementing the react-and-wind technique, but also to utilize REBCO CORC$^{\circledR}$ cable, whose rigidity makes small bending radius a possible source of conductor degradation. Similar to the block design, the common-coil allows aligning the block with the flux lines, thus minimizing the HTS conductor use. Also, by having the layer-to-layer splice inside the winding pole at the center of the coil, one can wind and react individual layers, and \say{grade} each layer to maximize efficiency.  The two top designs in Fig. \ref{fig:20T-cross-sections} use Bi2212 cables in the small blocks around the aperture and in the first layer, followed by either 3 or 5 layers of \nbsn cables. In both designs, the coil area and width are small compared to the previous designs; however, it is important to point out that the load-line margins are below the 15\% criteria. In the third design, we implement a large CORC$^{\circledR}$ wire, in series with 4 layers of HTS. Also in this case further magnetic analysis will be carried out to bring the load-line margin to the design criteria. Regarding the coil stress, the common coil design allows the insertion of vertical plates to intercept the horizontal e.m. forces, and as in the BL design, horizontal bars can be used to intercept the vertical force. As a next step, a mechanical analysis will be performed to verify the stress, and the magnetic design will be updated accordingly.

\paragraph{Conclusions - 20T design concepts}
We presented in this section a preliminary investigation of a hybrid 20 T dipole, which we consider a promising option for a dipole magnet operating beyond the limits of Nb$_3$Sn and aimed at minimizing the HTS volume. Two HTS conductors are considered: Bi2212, in the form of a Rutherford cable with $J_e$ of $740$ A/mm$^2$ at $20$ T, and REBCO tape in a CORC$^{\circledR}$/STAR$^{\circledR}$ wire with $J_e$ of $590$ A/mm$^2$ at $20$ T. As part of the design criteria, we target a bore field of $20$ T with a load-line margin of at least 15\% for both LTS and HTS coils. Also, all the coils are assumed powered in series, and stress must be kept below $150-180$ MPa in the \nbsn and below $120$ MPa in the HTS. A preliminary analysis done with sector coils indicated that 1) with identical $J_0$ in both HTS and LTS, we have a coil width of $\sim 70$ mm, 2) radial stresses of about $200$ MPa are generated by the radial/horizontal e.m. forces, and 3) a significant reduction of HTS area can be obtained by \say{anti-grading}, i.e. by increasing the size of the \nbsn outsert. Finally, we performed a preliminary analysis of a $20$ T hybrid with different coil design options, all shown in Fig. \ref{fig:20T-cross-sections}. The designs are not yet comparable since they do not meet all the specifications, but they provide a first idea of the overall coil features, and they constitute a starting point for further analysis. 
\begin{tcolorbox}[colback=yellow!10!white,
                     colframe=black!75,
                     coltitle=black,
                     colbacktitle=black!10,
                     fonttitle=\bfseries,
                     title=Statement \thestatements: 20 T accelerator dipoles \textit{may} be within reach,  
                     center, 
                     valign=top, 
                     halign=left,
                     before skip=0.8cm, 
                     after skip=1.2cm,
                     center title, 
                     width=15cm]
 
     Judicious magnet design, incorporating concepts such as stress management and optimized pre-stressing, indicates 20 T dipole fields may be achievable with existing superconducting materials in a variety of magnet configurations. 
  \end{tcolorbox}
  \stepcounter{statements}

\subsection{Technology underpinning the development of advanced magnet technology}
There are many elements of technology development that lie at the heart of advanced magnet \randd. Advances in modeling are providing more detailed insight into stress and strain states in magnets, including full 3D effects and taking into account complex nonlinear material and interface behavior. We are now seeing synergies with the broader Accelerator Modeling community (see associated whitepaper \cite{biedron11}). Advances in diagnostics provide new and unique insight into the timescales and energy scales of perturbations that may lead to quenches and to magnet training, and to their location in the magnet. Advances in materials and testing techniques may lead to reduction in training. These elements are a significant part of the US MDP and provide lasting progress for the broader superconducting magnet community. Campanion whitepapers that explore these areas in more detail include \cite{baldinifiber}, \cite{marchevskydiagnostics},  \cite{marchevskytraining}, and \cite{stoynev}. 

Advancing magnet technology requires a detailed understanding of magnet performance and insight into magnet behavior and performance limitations through optimized diagnostics. A key element of the US MDP approach is the development and implementation of an integrated diagnostics package of novel sensor hardware, electronics and data analysis techniques for real-time, non-invasive monitoring of LTS, HTS and hybrid magnets. This entails synchronous acquisition of voltages, acoustic, magnetic and optical data for magnets under test, and a synergistic data analysis. The end goal of such development would be the application of this advanced diagnostic system to existing and future accelerator magnets and magnet test facilities. Current MDP  \randd elements for superconducting magnet diagnostics include the following:
\begin{itemize}
\item	Develop a next-generation acoustic emission diagnostic hardware capable of self-calibration to drastically improve disturbance triangulation accuracy and \say{fingerprinting}. Use it to study physics of quench-triggering disturbances and mechanisms of mechanical memory and training in \nbsn magnets.
\item	Establish fiber-optic based diagnostic capabilities through the use of Fiber Bragg Grating (FBG) and Rayleigh scattering-based sensors to measure elastic deformations, localize hot spots (especially in HTS magnets) and probe mechanical disturbances in SC cables.
\item	Improve accuracy of voltage, magnetic and acoustic-based diagnostics through calibration using distributed spot heater and piezo-transducer arrays.
\item	Bring magnetic diagnostics to the next level through development and use of flexible multi-element quench antennas, large-scale Hall sensor arrays and non-rotating field quality probes, aiming at understanding electromagnetic instabilities in LTS magnets and imaging current-sharing patterns in superconducting cables and HTS magnet coils. Develop new algorithms for current flow reconstruction and disturbance localizations.
\item	Design and conduct innovative small-scale experiments to probe training behavior and energy release in impregnated cables under similar loads as in the magnets. 
\item	Develop new methods for reliable and robust quench detection and localization for HTS magnets and hybrid LTS/HTS magnets.
\item	Use diffuse field ultrasonic techniques to enable targeted delivery of vibrational excitation to the conductor, for a non-invasive structural local probing of SC coils and mitigation of their training behavior.
\item	Apply machine learning and deep learning approaches to process diagnostic data and identify real-time predictors of magnet quenching
\item	Develop cryogenic digital and analog electronics to facilitate, simplify and improve reliability of diagnostic instrumentation by enabling pre-processing of magnet diagnostic data in the cryogenic environment.
\end{itemize}
A synergistic analysis of data acquired by these diverse diagnostic techniques will bring us closer to answering key technical questions that define SC magnet performance. It is an ample and comprehensive program with the aim of developing an integrated system of hardware and software solutions applicable not only to the U.S. MDP SC magnets, but to any other SC accelerator magnet and also magnets for test facilities. The effort should extend well beyond MDP and engage instrumentation experts across national and international labs. 

In addition to the diagnostics that provide critical insight into magnet behavior, a number of potential techniques aiming to affect the training rate before or during magnet powering are under investigation:
\begin{itemize} 
\item High-Cp conductor and insulation development that can lead to conductors and coils with optimized characteristics enabling stable operation against perturbations \cite{Ligfa}, \cite{barzi1};
\item Artificially increasing the coil current during a quench by discharging a large capacitor at quench detection. The \say{overcurrent} transient will generate forces beyond the nominal quench current, possibly significantly increasing the rate at which the magnet trains and
hence reducing the total number of training quenches;
\item Inducing ultrasonic vibrations into the conductor and coil parts during ramp up. This may
allow for gradual energy discharges, avoiding accumulation of energy (notably due to friction) in areas around the coil/conductor and potentially improving the rate at which magnets train;
\item  Performing fast-turn-around cable/stack training \randd in a controllable configuration, thereby providing more quantitative insight into training mechanisms and means to influence them.
\end{itemize} 

These approaches complement other technology areas that are focused on understanding the disturbance spectrum and on mitigating their sources.
\begin{tcolorbox}[colback=yellow!10!white,
                     colframe=black!75,
                     coltitle=black,
                     colbacktitle=black!10,
                     fonttitle=\bfseries,
                     title=Statement \thestatements: Advanced diagnostics provide "science" of magnets,  
                     center, 
                     valign=top, 
                     halign=left,
                     before skip=0.8cm, 
                     after skip=1.2cm,
                     center title, 
                     width=15cm]
     New diagnostics and testing techniques are providing extraordinary insight into magnet behavior and provide critical feedback to conductor and magnet designers. The techniques and knowledge gained are valuable to the broader superconducting magnet community.
  \end{tcolorbox}
  \stepcounter{statements}

\section{Nurturing a strong industrial magnet ecosystem for science and society}
Science facilities, particularly those of the scale of a future collider for high energy physics, depend fundamentally on industry. A strong industrial capacity can be very effective at reducing cost of components, and enables timely delivery to meet project schedules. 

The HEP community recognizes that the timescales for the germination of a future collider project are inconsistent with the timescales of industry; hence an approach that  focuses on nurturing an \say{ecosystem} of industrial capability and expertise, rather than on industrial capacity, is more viable between HEP projects.

\subsection{Building and leveraging capabilities within US industry }
Nurturing the industrial ecosystem can, and should, be pursued on multiple fronts:
\begin{itemize}
    \item Support technology transfer and public private partnerships related to HEP technology, e.g. advanced magnet systems, that have potential to seed new applications in society. Examples include medical gantries, compact fusion, superconducting magnets for wind generation, etc. An excellent example of a DOE program that targets this avenue to nurturing the ecosystem is the DOE Accelerator Stewardship program.
    \item Align DOE investments in small businesses with HEP community needs. The DOE SBIR program strives to nurture the industry ecosystem, and HEP programs, such as the MDP, can work to encourage industry involvement in the program around topics and capabilities that support future collider technologies.
    \item The industry ecosystem depends on a pipeline of talent and expertise. HEP can work with labs and Universities to make sure a steady stream of scientists and engineers are produced that can support industry needs.
\end{itemize}

The success of the DOE-OHEP Accelerator Stewardship Program, initiated following the report \say{Accelerators for America’s Future} \cite{accel1}, catalyzed the creation of a new program office in DOE, the Office of Accelerator \randd and Production (ARDAP). Its mission is to \textit{\say{...help coordinate Office of Science (SC) accelerator R\&D, foster public-private partnerships to develop and deploy accelerator technology, support workforce development and improve its diversity, and provide resources for accelerator design and engineering. The overarching goal is to ensure a robust pipeline of innovative accelerator technology, train an expert workforce, and reduce significant supply chain risks by re-shoring critical accelerator technology.}} ARDAP is therefore ideally suited to support the industrial ecosystem described above, that is so essential to deliver on future collider opportunities.

\begin{tcolorbox}[colback=yellow!10!white,
                     colframe=black!75,
                     coltitle=black,
                     colbacktitle=black!10,
                     fonttitle=\bfseries,
                     title=Statement \thestatements: Work with ARDAP to strengthen the magnet ecosystem,  
                     center, 
                     valign=top, 
                     halign=left,
                     before skip=0.8cm, 
                     after skip=1.2cm,
                     center title, 
                     width=15cm]
     The newly minted Office of Accelerator \randd and Production (ARDAP) is ideally suited to work with, and support, DOE-OHEP needs in terms of domestic industrial expertise and capabilities for future colliders. MDP can support HEP by clearly identifying critical needs, both near  and long-term, for HEP accelerator magnet development and production.
  \end{tcolorbox}
 \stepcounter{statements}

\section{Opportunities and challenges for international collaboration}
High Energy Physics is fundamentally an international endeavor, and the development of critical technologies enabling new colliders is similarly advanced through the combined efforts of international scale. All areas of the MDP benefit from close communication and collaboration with magnet and conductor \randd partners outside the US. Well-established communication channels ensure that we maximize progress through programs that are both competitive but also complementary.

Ties with international laboratories and universities have been, and we expect will continue to be, a critical element of our program. The US MDP leadership strives to foster a diverse, inclusive, and innovative culture that motivates innovation and open communication. Transparency in our purpose and research approach has proven to be effective in supporting the development of strong collaborations with international partners.

Furthermore, we note that a significant fraction of MDP staff were educated, wholly or partially, at international institutions. Exchanges of ideas, concepts, and results with international collaborators is central to our approach to research. 

The European Strategy process has recently been completed, resulting in a suite of recommendations related to investments in critical technologies for future colliders. The implementation of those recommendations is currently underway; the European Laboratory Directors Group (LDG) initiated the development of roadmaps for each area, including magnet technology - the European High Field Magnet program (HFM) \cite{EU22}, \cite{vedrinehfm}. That process is now completed, and teams from multiple European laboratories and Universities are currently organizing to deliver on the European magnet roadmap. 

 The US MDP was actively engaged in the European roadmapping process, and we now have the opportunity to collaborate with the HFM through our established programs. A major consideration is to document collaboration topics to make sure efforts are equitable, and to insist on transparency and openness among the teams, while encouraging innovation and recognizing individual's contributions to the programs.
 
\begin{tcolorbox}[colback=yellow!10!white,
                     colframe=black!75,
                     coltitle=black,
                     colbacktitle=black!10,
                     fonttitle=\bfseries,
                     title=Statement \thestatements: High Energy Physics is an international endeavor, 
                     valign=top, 
                     halign=left,
                     before skip=0.8cm, 
                     after skip=1.2cm,
                     center title, 
                     width=15cm]
     High Energy Physics is an international endeavor, and magnet technology research for the field is accordingly international in nature. Equity and openness are sought with collaborators with common interests and relevant expertise and capabilities.
  \end{tcolorbox}
 \stepcounter{statements}
 
\section{Summary}
High field magnet technology is essential to the mission of High Energy Physics. A robust long range \randd program serves as an investment in the field's future, providing lasting advances and understanding in magnet performance, with innumerable benefits to the broader superconducting magnet community. Furthermore, such a longstanding \randd effort serves to lead and align industry, in particular small businesses, with the field, and potentially to initiate and incubate new applications of superconducting technology for society.

A long range \randd magnet program, i.e. the MDP, is a highly leveraged investment that provides advanced technology that paves the way for future collider projects, providing a solid foundation of knowledge and expertise for directed \randd programs when opportunities arise. We note that the MDP serves as an excellent venue for nurturing new talent and maintaining expertise that is critical to HEP's future; the program is highly sought after by early career scientists, and is designed and led with the intention of growing the next generation of magnet scientists, providing continuity and promise for our field.

\newpage

\printbibliography
\end{document}